\documentclass[aps,twocolumn,amsmath,amssymb,amsthm,dsfont,prb,nofootinbib]{revtex4-1}
\usepackage{times}
\usepackage{graphicx}
\usepackage{mathrsfs}
\usepackage{dcolumn}
\usepackage{bm}
\usepackage[colorlinks=true,citecolor=blue,linkcolor=blue]{hyperref}
\usepackage{tikz}
\usepackage{hhline}
\usepackage{float}
\usepackage{scalerel}
\usepackage{amsfonts}
\usepackage{bbm}
\usepackage{tabularx}
\usepackage{multirow}
\usepackage{tikz-cd}
\usepackage{amsthm}
\usepackage{amsbsy}
\usepackage[T1]{fontenc}
\usepackage{ mathdots }
\usepackage[caption=false]{subfig}
\usepackage{colortbl}
\usepackage{tabu}
\usepackage{xcolor}
\usepackage{enumitem}

\makeatletter
\def\l@subsection#1#2{}
\def\l@subsubsection#1#2{}
\makeatother

\makeatletter
\renewcommand*\env@matrix[1][*\c@MaxMatrixCols c]{%
  \hskip -\arraycolsep
  \let\@ifnextchar\new@ifnextchar
  \array{#1}}
\makeatother

\newtheorem{thm}{Theorem}
\newtheorem{lemma}[thm]{Lemma}
\newtheorem{prop}[thm]{Proposition}

\DeclareMathOperator{\im}{im}

\begin{document}

\def\bra#1{\mathinner{\langle{#1}|}}
\def\ket#1{\mathinner{|{#1}\rangle}}
\def\inner#1{\mathinner{\langle{#1}\rangle}}
\def\bs#1{\boldsymbol{#1}}
\def\tbs#1{\tilde{\boldsymbol{#1}}}
\newcommand{\Ket}[1]{\vcenter{\hbox{$\displaystyle\stretchleftright{|}{#1}{\bigg\rangle}$}}}

\def\linecolor{red}
\def\lc{\taburulecolor{\linecolor}} 

\def\includegraphicsr#1#2{\raisebox{-.5\height}{\includegraphics[scale={#1}]{#2}}}

\title{Jordan-Wigner Dualities for Translation-Invariant Hamiltonians in Any Dimension: Emergent Fermions in Fracton Topological Order}

\author{Nathanan Tantivasadakarn}
\affiliation{Department of Physics, Harvard University, Cambridge, MA 02138, USA}

\begin{abstract}
Inspired by recent developments generalizing Jordan-Wigner dualities to higher dimensions, we develop a framework of such dualities using an algebraic formalism for translation-invariant Hamiltonians proposed by Haah. We prove that given a translation-invariant fermionic system with general $q$-body interactions, where $q$ is even, a local mapping preserving global fermion parity to a dual Pauli spin model exists and is unique up to a choice of basis. Furthermore, the dual spin model is constructive, and we present various examples of these dualities. As an application, we bosonize fermionic systems where free-fermion hopping terms are absent ($q \ge 4$) and fermion parity is conserved on submanifolds such as higher-form, line, planar or fractal symmetry. For some cases in 3+1D, bosonizing such a system can give rise to fracton models where the emergent particles are immobile but yet can behave in certain ways like fermions. These models may be examples of new nonrelativistic 't Hooft anomalies. Furthermore, fermionic subsystem symmetries are also present in various Majorana stabilizer codes, such as the color code or the checkerboard model, and we give examples where their duals are cluster states or new fracton models distinct from their doubled CSS codes.
\end{abstract}

\maketitle
\tableofcontents

\section{Introduction}\label{intro}
The Kramers-Wannier (KW) and Jordan-Wigner (JW) transformations are important dualities in one-dimensional systems; a one-dimensional transverse-field Ising model with global $\mathbb Z_2$ spin-flip symmetry is KW dual to itself with an inverse Ising coupling. At the same time, it is also JW dual to a spinless-fermionic system with global $\mathbb Z_2^F$ fermion parity symmetry. 

There are generalizations of both KW and JW dualities, though those of the former are much better understood.  KW dualities have been extended to transverse-field Ising models in arbitrary dimensions\cite{Wegner1971,Kogut1979}, which gives rise to dual gauge theories whose ground states can exhibit topological order\cite{Kitaev2003,BombinMartin-Delgado2008}. At the same time, generalized Ising models with multibody Ising interactions can exhibit extra symmetries in addition to the usual global spin-flip symmetry, such as higher-form symmetries\cite{Gaiottoetal2015,KapustinThorngren2017,Yoshida2016,TsuiWen2020} or subsystem symmetries\cite{NewmanMoore1999,XuMoore2004,VijayHaahFu2016,Youetal2018,Devakuletal2018}. These exotic symmetries preserve the Hamiltonian under a spin flip on individual subdimensional manifolds, with integer or possibly fractal dimensions. KW dualities have been generalized to theories with such symmetries\cite{CobaneraOrtizNussinov2011,VijayHaahFu2016,Williamson2016,KubicaYoshida2018,Pretko2018,ShirleySlagleChen2019,Radicevic2019}, and the dual theories can exhibit new types of ordered states, such as fracton topological order\cite{Haah2011,Yoshida2013,VijayHaahFu2016}.

JW dualities can also be generalized to higher dimensions, though the intricacies have only been fully understood recently. The parallel to KW dualities has been established as an exact bosonization introduced in Refs. \onlinecite{ChenKapustinRadicevic2018,ChenKapustin2019,Chen2019} (see also, Ref. \onlinecite{BravyiKitaev2002}) in the case of a global fermion parity symmetry. However, the construction is more subtle than its bosonic counterpart. The duality is only well defined if the fermion theory is put on a spin manifold and explicitly depends on a choice of spin structure. Furthermore, for spatial dimension $d>1$, the resulting gauge theory is different from the usual gauge theory obtained by a KW duality. The gauge constraints can be understood as an anomalous higher-form symmetry that is required so that a spin system can support emergent excitations with fermionic statistics\cite{GaiottoKapustin2016}.

\begin{figure}[t!]
\centering
    \begin{subfloat}
        \centering
       \raisebox{-0.5\height}{\includegraphics[scale=0.55]{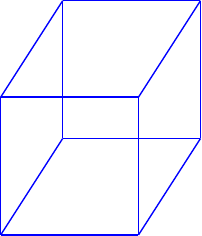}}
    \end{subfloat}
    \begin{subfloat}
        \centering
        \raisebox{-0.5\height}{\includegraphics[scale=0.55]{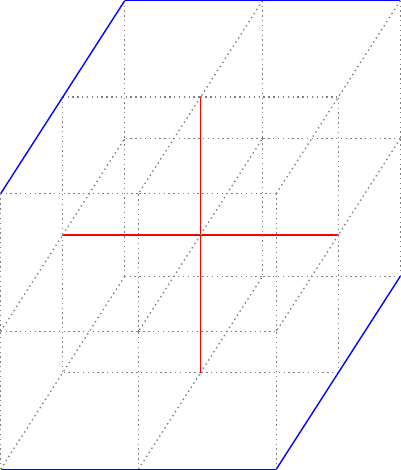}}
    \end{subfloat}
        \begin{subfloat}
        \centering
        \raisebox{-0.5\height}{\includegraphics[scale=0.55]{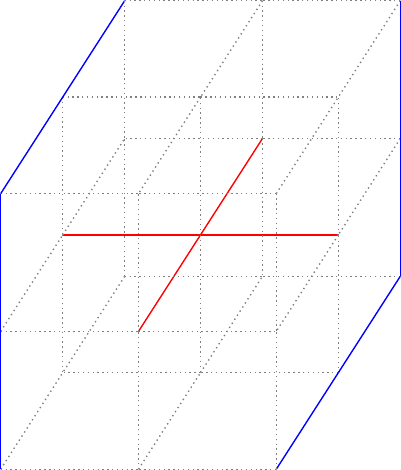}}
    \end{subfloat}
        \begin{subfloat}
        \centering
        \raisebox{-0.5\height}{\includegraphics[scale=0.55]{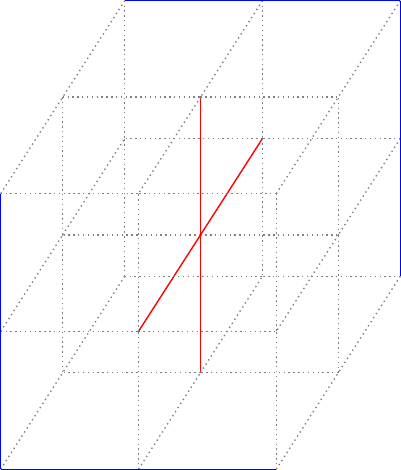}}
    \end{subfloat}
    \caption{An example of an exactly solvable model with fracton excitations that are fermions, which is a twisted version of the X-Cube stabilizer code. The blue and red lines respectively denote Pauli $Z$ and $X$'s living on the edges of a cubic lattice. The model can be thought of as the result of gauging a fermion system which conserves fermion parity in the (001), (010) and (001) planes of a dual cubic lattice. We discuss how to obtain this model in Sec. \ref{ex:planar3D_1}.}
    \label{fig:twistedXcube}
\end{figure}

It is natural to ask whether there is a similar analog to the KW dualities between generalized Ising models with many-body Ising interactions. That is, whether one can generalize JW dualities to fermionic systems with arbitrary $q$-body interactions (where $q$ is even), which conserve fermion parity on individual subdimensional manifolds. In this paper, we answer this question in the positive and explicitly construct an exact bosonization for any such interaction to spin systems with unusual (and possibly anomalous) symmetries assuming translation invariance. The key step in establishing such a duality takes advantage of an algebraic representation of translation invariant Pauli and Majorana Hamiltonians introduced in Refs. \onlinecite{Haah2013} and \onlinecite{VijayHaahFu2015}, respectively. Furthermore, the constructed model can be shown to be unique up to a choice of basis. More specifically, all possible dual models only differ by a finite-depth translation-invariant Clifford circuit.

Interestingly, for certain types of fermion parity symmetries in 3D, our generalized JW duality allows us to construct gauge theories whose excitations are immobile in the deconfined phase, that is, the ground state exhibits fracton order. However, such excitations are also fermions, in the sense that they cannot be condensed due to the noncommutativity of the operators that proliferate them. Moreover, it is understood that the fermions can only be condensed if paired up with another excitation (either physical or emergent) exhibiting the same anomaly\cite{GaiottoKapustin2016,AasenLakeWalker2019}. An example of a model which exhibits such property is a twisted X-cube model shown in Fig. \ref{fig:twistedXcube}. This opens a question of whether there are meaningful statistical processes that one can perform to detect whether such immobile excitations are fermionic. A closely related question is whether the gauge constraints of these fracton models, when considered as a higher-rank symmetry\cite{Pretko2017,MaHermeleChen2018,BulmashBarkeshli2018,Gromov2019,Seiberg2019,WangXuYau2019}, has an associated 't Hooft anomaly that generalize those discussed in Ref. \onlinecite{GaiottoKapustin2016}.

An important application of these dualities (and its predecessors) is that they are local maps, and are hence useful for simulating arbitrary interacting fermionic systems in any dimension with qubits. For example, any translation-invariant Majorana code \cite{BravyiTerhalLeemhuis2010,VijayHsiehFu2015,VijayHaahFu2015} can be locally mapped to a translation-invariant Pauli stabilizer code, and any operations used to perform the computation also map accordingly\footnote{Note that our mapping is different from doubling a Majorana code into a self-dual CSS code}.

\begin{table*}[!t]
\caption{Summary of JW Dualities considered, and the corresponding ground states. ?? means the dual bosonic ground state has not been identified. The subscript $F$ denotes a twisted version where emergent excitations are fermions. TC, TO, and SSB stand for Toric Code, Topological Order, and Spontaneous Symmetry Breaking, respectively.}
\begin{tabular}{|c|c| c|c|c | c|c|}
\hline
Sec. &$d$ & $\mathbb Z_2^F$ &    $\mathbb Z_2$  & $\mathbb Z_2$ anomalous  & Fermionic ground state & Bosonic ground state\footnote{Here, the ground state refers to that of the dual spin Hamiltonian along with locally-generated symmetry constraints imposed either as a strict gauge constraint or enforced energetically.}\\
\hline
\ref{ex:Global2D}&2 & Global & 1-form & Yes & Product & 2D TC\cite{ChenKapustinRadicevic2018}\\
&&&&& Majorana color code\cite{VijayHsiehFu2015} & $\mathbb Z_2^2$ TO\\
  \hline
\ref{ex:Global3D}&3 & Global & 2-form & Yes & Product & 3D TC$^F$\cite{ChenKapustin2019}\\
&&&&& Majorana checkerboard\cite{VijayHaahFu2015} & Semionic X-cube\cite{MaLakeChenHermele2017} $\otimes$ 3D TC$^{F}$? \\
&&&&& Majorana codes 2-5\cite{VijayHaahFu2015} & ??\\
\hline
\ref{ex:1form}& 3 & 1-form & 1-form & Yes? & Product & 3D TC\\
\hline
\ref{ex:line2D_1}& 2 & Line$(\times 2)$ & Line $(\times 2)$ & No & Product & SSB\cite{XuMoore2004}\\
&&&&& SSB$_y$& Product\\
&&&&& SSB$_x$ & SSPT\cite{Youetal2018,DevakulWilliamsonYou2018}\\
&&&&& $\mathbb Z_2$ TO  & Wen plaquette\cite{Wen2003}\\
\hline
\ref{ex:2DSSPT2}& 2 & Line $(\times 2)$ & Line$(\times 2)$ & No & Product & SSB\\
&&&&& SSB$_x$& Product\\
&&&&& SSB$_y$ & SSPT\cite{DevakulWilliamsonYou2018}\\
&&&&& Majorana color code  & SSPT\cite{DevakulWilliamsonYou2018}\\
\hline
\ref{ex:planar3D_1}& 3 & Planar ($\times 3$) &  Rank-2 & Yes? & Product & X-cube$^{(F?)}$\\
&&&&& Majorana code 3 & ??\\
\hline
\ref{ex:planar3D_2}& 3 & Planar ($\times 6$) &  Planar ($\times 6$) & No & Product & SSB\\
&&&&& Majorana Checkerboard & 3D Cluster state\\
\hline
\ref{ex:Fibonacci2D}& 2 & 2D Fractal  & 2D Fractal & Yes & Product & SSB\\
\hline
\ref{ex:Fibonacci3D}& 3 & 2D Fractal stacks & Higher-rank fractal & Yes & Product & Yoshida's fractal code$^F$\\
\hline
\ref{ex:Haah}& 3 & 3D Fractal & Higher-rank fractal & Yes? & Product & Haah's code$^{(F?)}$\\
&&&&& Majorana code 5 & ??\\
\hline
\ref{twistedCSS}& 3 & Planar ($\times 3$) & Rank-2 & Yes? & Product & Checkerboard$^{(F?)}$\\
&&&&& Majorana Checkerboard & ??\\
\hline
\end{tabular}
\label{tab:resultssummary}
\end{table*}

This paper is structured as follows: In Sec. \ref{review}, we review the KW and JW dualities on a 2D square lattice. In Sec. \ref{KWduality}, we introduce the algebraic formalism of translation-invariant Pauli Hamiltonians and use it to construct a KW duality between generalized transverse-field Ising models in any dimension. In close parallel, Sec. \ref{JWduality} constructs the generalized JW duality for a fermion model with arbitrary $q$-body interactions, where we prove its existence and uniqueness up to a choice of basis. Sec. \ref{Examples} discuss various examples, from reviewing the dualities with global fermion parity in 2D and 3D in this formalism, to new dualities where the fermionic system has additional higher-form or subsystem fermion parity symmetry. A summary of the dualities and models considered in this section are summarized in Table  \ref{tab:resultssummary}. Sec. \ref{twistedCSS} outlines a general procedure to construct twisted versions of translation-invariant CSS codes. Readers interested in fracton models can directly go to Secs., \ref{ex:planar3D_1}, \ref{ex:Fibonacci3D}, \ref{ex:Haah}, and \ref{twistedCSS}. In Sec. \ref{'tHooft}, we conjecture and give supporting arguments that for certain JW dualities, the gauge constraints of the dual spin models exhibit an anomaly associated to the fact that the emergent particles are fermions. We also give a concrete example where an anomalous fractal symmetry can be realized as an effective symmetry action on the boundary of a bulk Symmetry-Protected Topological (SPT) phase.  We conclude in Sec. \ref{Discussion} with various open questions.

\section{Review of KW and JW dualities in 2D}\label{review}
Before diving into the algebraic formalism, we find it helpful to review the KW and JW dualities on a square lattice in the standard notation.
\subsection{KW duality in 2D}\label{KWreview}
We consider a square lattice where qubits are placed on each vertex. The lattice can be infinite or can have periodic boundary conditions with a very large system size. The Hamiltonian is given by a transverse-field Ising model
\begin{align}
H&= -\sum_{<ij>} Z_iZ_j -h\sum_i X_i \nonumber\\
 &= -\sum_v  \left [ \includegraphicsr{1}{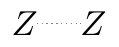} +  \includegraphicsr{1}{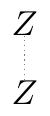} + h \includegraphicsr{1}{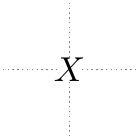}  \right ]
    \label{equ:2DTFIM}.
\end{align}
where $h$ is the strength of the transverse field.

The KW duality is an isomorphism between the algebra of the operators in this Hamiltonian and the algebra of operators of a ``dual'' (generalized) transverse-field Ising model, where the roles of the Ising term and transverse fields are swapped. Without loss of generality, we will choose a basis where the transverse field points in the $x$ direction (and therefore represented by a Pauli $X$) and all the Ising terms commute. The mapping can be described roughly as follows
\begin{enumerate}
    \item For each transverse field $X$, locate the position of all the Ising terms which anticommute with it, and map $X$ to a product of Pauli $Z$'s at those positions.
    \item Map each Ising term to a transverse field $X$ living at that exact same position.
\end{enumerate}
In the case of the 2D Ising model above, the Ising term is defined for each edge of the square lattice. Therefore, we map each Ising term to a Pauli $X$ on that edge, and map the Pauli $X$ on each vertex to a product of Pauli $Z$'s on edges emanating from that vertex. Pictorially,
\begin{align}
\includegraphicsr{1}{2DKW_Zx.pdf} &\leftrightarrow \includegraphicsr{1}{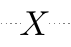},\\
\includegraphicsr{1}{2DKW_Zy.pdf} &\leftrightarrow \includegraphicsr{1}{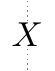},\\
    \includegraphicsr{1}{2DKW_X.pdf} &\leftrightarrow \includegraphicsr{1}{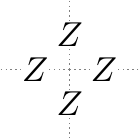},
\end{align}
where the original operators are on the left and the dual operators are on the right. We see that the commutation relations of all operators are preserved. Therefore, we have dualized the Hamiltonian to
\begin{align}
    \tilde H = -\sum_v  \left [\includegraphicsr{1}{2DKW_Xty.pdf} + \includegraphicsr{1}{2DKW_Xtx.pdf}+ h\includegraphicsr{1}{2DKW_Zt.pdf}  \right ].
    \label{equ:2DKWdualHam}
\end{align}
which is a generalized Ising model with a four-body Ising term. By rescaling the Hamiltonian, we see that the strength of the transverse field is $1/h$.

At the level of states, the KW duality is more subtle due to constraints arising from the duality. For example, the product of dual Ising terms over all vertices is the identity. Dualizing, we find that the product of Pauli $X$ on all vertices  in the original Ising model must also be identity
\begin{align}
    \prod_v \includegraphicsr{1}{2DKW_X.pdf} =1 &\leftrightarrow \prod_v \includegraphicsr{1}{2DKW_Zt.pdf} =1.
\end{align}
This constraint on the original Ising model is a symmetry constraint. Namely, it says that only states that are even under the global $\mathbb Z_2$ symmetry $\prod_v X_v$ are allowed to map under the duality.

Similarly, the product of Ising terms also produces symmetry constraints in the dual Ising model. The product of $\includegraphicsr{1}{2DKW_Zx.pdf}$ and $\includegraphicsr{1}{2DKW_Zy.pdf}$ around a plaquette, or in general around any cycle of the square lattice (regardless of contractibility), must be the identity. Therefore, in the dual Ising model, it enforces the following constraints
\begin{equation}
    \includegraphicsr{1}{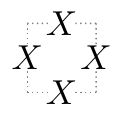}= \includegraphicsr{1}{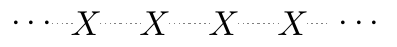}=\includegraphicsr{1}{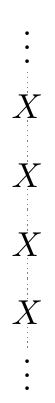}=1.
    \label{equ:1formsym}
\end{equation}
The symmetries enforced here perform a spin flip on a submanifold (i.e., the 1-cycles) of the square lattice and are called 1-form symmetries\cite{Gaiottoetal2015}.

Based on the discussion above, let us call the selection of operators that product to the identity \textit{identity generators} and the set of operators that commute with the transverse-field Ising Hamiltonian \textit{symmetries}. The example above shows that dualizing the identity generators result in the symmetries of the dual Hamiltonian. Indeed, this can be seen more generally from the fact that the product of the identity generators trivially commutes with all other operators. Hence, since the commutation relations are preserved, the dual operator must also commute with the dual Hamiltonian and is thus a symmetry. We will later show a stronger statement using the algebraic formalism that these identity generators are in one-to-one correspondence with the symmetry constraints of the dual theory and vice versa.

Lastly, we review how the phase diagram maps under the duality. When $h\ll 1$, the ground state of the Hamiltonian \eqref{equ:2DTFIM} is in the ferromagnetic (symmetry broken) phase, while when $h \gg 1$, the ground state is in the paramagnetic (symmetry preserving) phase. On the other hand, the ground state of the dual Hamiltonian \eqref{equ:2DKWdualHam} is more subtle without symmetry constraints. For $h\gg 1$, the ground state is not gapped because the vertex terms imposes at most $N$ constraints (where $N$ is the number of vertices), while the system consists of $2N$ sites. Nevertheless, the ground state can be gapped if the local symmetry constraint \includegraphicsr{1}{2DKW_1form1.pdf} in Eq. \eqref{equ:1formsym} is either imposed as a strict $\mathbb Z_2$ gauge constraint or added to the dual Hamiltonian so that it is enforced energetically. In either case, the resulting phase diagram is that $h\ll 1$ corresponds to the symmetric phase, while $h \gg 1$ corresponds to the symmetry broken phase under the 1-form symmetry. When thought of as a gauge theory, the two regimes are also known as confined and deconfined phases, respectively.

Those familiar with the toric code will recognize this Gauss law and the Ising term of this 1-form symmetry as the plaquette and vertex terms, respectively. However, it is important to emphasize that the regime where $h \gg 1$ does not imply a duality between a paramagnet Hamiltonian and the toric code Hamiltonian, since they have different ground state degeneracies on the torus. It is only when the noncontractible symmetry constraints in Eq. \ref{equ:1formsym} (which can be thought of as the Wilson loops) are also strictly imposed that the dual Hamiltonian has a unique ground state, allowing the duality to be possible. To properly obtain a duality to full the toric code Hamiltonian, which is a dynamical $\mathbb Z_2$ gauge theory, one must properly couple the transverse-field Ising model to an additional dynamical gauge field living on each edge of the lattice. We refer to Ref. \onlinecite{Radicevic2018} for a thorough treatment of such a duality.

\subsection{JW duality in 2D}\label{JWreview}
We will now perform a similar exercise to obtain the JW duality in 2D, which was introduced in Ref. \onlinecite{ChenKapustinRadicevic2018}. We consider a complex fermion living on each vertex of the square lattice. The complex fermion operator $c_v$ at each vertex can be decomposed into two real (Majorana) fermions
\begin{align}
\gamma_v  &= c_v+c_v^\dagger,  & \gamma_v' = (c_v-c_v^\dagger)/i,
\end{align}
so that the local fermion parity operator is
\begin{align}
P_v = 1-2c_v^\dagger c_v = -i\gamma_v \gamma_v'.
\end{align}
We consider a fermionic Hamiltonian
\begin{align}
    H = -\sum_v  \left [\includegraphicsr{1}{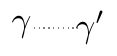} + \includegraphicsr{1}{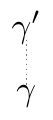} +\mu \includegraphicsr{1}{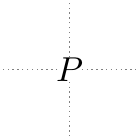} \right ],
    \label{equ:2Dglobalfermion}
\end{align}
where the Ising terms have been replaced by fermionic hopping operators and the transverse field is now the fermion parity operator with field strength $h$ replaced by the chemical potential $\mu$. We note that it is possible to add terms with next-nearest neighbor hoppings and further interactions but since they can all be written as products of these three operators, it is straightforward to generalize.

The JW duality is an isomorphism between the (parity even) algebra of fermions to the algebra of operators in a dual Pauli Hamiltonian. In analogy to the prescription of the KW duality previously, the procedure is given descriptively by the following
\begin{enumerate}
    \item For each fermion parity operator $P$, locate the position of all hopping terms which anticommute with it and map $P$ to a product of Pauli $Z$'s at those positions.
    \item Map each hopping term to $X$ living at that exact same position, \textit{accompanied by additional Pauli $Z$'s at certain positions to preserve the commutation relations}.
\end{enumerate}
The italicized part of step 2 is for now perhaps vague, since we have not yet specified how to determine such additional positions. We will later make this step precise using the algebraic formalism and also prove that such a choice always exists (see Lemma \ref{prop:Texists}). For now, let us see how this works out in the 2D example. The duality can be chosen to be
\begin{align}
\raisebox{-\height}{\includegraphics{2DJW_Sx.pdf}} &\leftrightarrow \includegraphicsr{1}{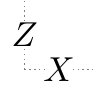},\\
\includegraphicsr{1}{2DJW_Sy.pdf} &\leftrightarrow \includegraphicsr{1}{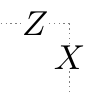},\\
    \includegraphicsr{1}{2DJW_P.pdf} &\leftrightarrow \includegraphicsr{1}{2DKW_Zt.pdf},
\end{align}
where in comparison to the dual Ising model previously, there are extra Pauli $Z$'s attached to each $X$ to ensure that the commutation relations are preserved. Therefore, the dual Hamiltonian is given by
\begin{align}
    \tilde H = -\sum_v  \left [  \includegraphicsr{1}{2DJW_Xtx.pdf} + \includegraphicsr{1}{2DJW_Xty.pdf}+\mu \includegraphicsr{1}{2DKW_Zt.pdf} \right ].
    \label{equ:JWdual2D}
\end{align}

Similarly in the JW duality, identity generators are in one-to-one correspondence with the dual symmetries. The product of \includegraphicsr{1}{2DKW_Zt.pdf} on all vertices is the identity, enforcing a global fermion parity symmetry $\prod_v P_v$ on the fermions. On the other hand, a certain product of a fermionic hopping operators and fermion parity operators around a cycle that is the identity generates the following three symmetry constraints in the dual Pauli Hamiltonian.
\begin{equation}
    \includegraphicsr{1}{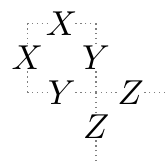}= \includegraphicsr{1}{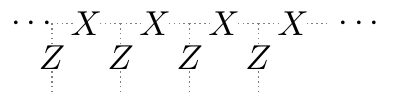}=\includegraphicsr{1}{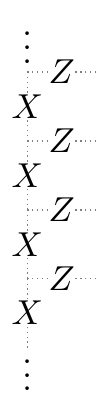}=1.
    \label{equ:1formF2D}
\end{equation}
Any other symmetry constraint can be written as a product of these three constraints and their translations.

The locally generated symmetry can be seen as a product of a plaquette term and a vertex term of the toric code. It therefore enforces an unusual Gauss law that requires charges and fluxes to be created in pairs, whose bound state are emergent fermions. The nonlocal symmetries can also be thought of as an anomalous 1-form symmetry corresponding to a closed string of the emergent fermions\cite{GaiottoKapustin2016}.

\section{KW Dualities in the Algebraic Formalism}\label{KWduality}
We will now review the algebraic formalism in the case of Pauli operators and see how the KW duality described previously fits into this language. We start by introducing the algebraic formalism in Sec. \ref{polynomialboson}. In Sec. \ref{transverseIsing}, we consider a generalized Ising model in this formalism, and show how to dualize it into another generalized Ising model where the roles of transverse field and Ising interactions are swapped. Along the way, we demonstrate the KW duality for the 1D transverse-field Ising model and repeat the 2D example in Sec. \ref{KW2D}.

\subsection{Algebraic Formalism}\label{polynomialboson}
The algebraic formalism of translation invariant Pauli Hamiltonians was first formulated in Ref. \onlinecite{Haah2013}. We remark that although the formalism was originally tailored to stabilizer codes (i.e., where all terms commute), it is still applicable to Pauli Hamiltonians whose terms do not necessarily commute, as we will explain below.
\subsubsection{Polynomial Representation of Pauli Operators}
Our system is a $d$-dimensional cubic lattice with qubits attached to each vertex. Fixing a point of origin, the position of each qubit can be labeled by its coordinates $\bs i =(i_1,...,i_d) \in \mathbb Z^d$.

Our goal is to represent all Pauli operators in this Hilbert space using a polynomial representation. A Pauli operator $p$ can be uniquely decomposed up to a phase as
\begin{equation}
    p \propto \bigotimes_{\bs i} Z_{\bs i}^{a_{\bs i}} X_{\bs i}^{b_{\bs i}},
    \label{equ:Pauliform}
\end{equation} 
where $a_{\bs i},b_{\bs i}\in\{0,1\}$ denotes the absence or presence of the Pauli $Z$ and $X$ at each position $\bs i$, respectively.  The position of Pauli $Y$'s are automatically included as the positions where both Pauli $Z$'s and Pauli $X$'s are present. We will algebraically represent this Pauli operator as a 2-component vector whose entries are polynomials
\begin{equation}
\bs{p}=\lc
 \begin{pmatrix}
    \sum_{\bs i}  a_{\bs i}  x_1^{i_1} \cdots x_{d}^{i_d} \\ \hline \sum_{\bs i}  b_{\bs i}  x_1^{i_1} \cdots x_{d}^{i_d}\end{pmatrix}.
    \label{equ:polynomialform}
\end{equation}
To unpack this notation, we sum over all monomials $x_1^{i_1} \cdots x_{d}^{i_d}$ for each Pauli operator that is present at coordinate $\bs i$. Furthermore, the red horizontal line has been drawn to visually separate the polynomial denoting positions of Pauli $Z$'s from that of Pauli $X$'s, and should not be confused with a fraction.

Let us give some examples. The Pauli $X_0$, $Y_0$, and $Z_0$ located at the origin can respectively be represented as vectors
\begin{align}
\lc
    \begin{pmatrix}
    0\\ \hline 1
    \end{pmatrix}
    , \begin{pmatrix}
    1\\ \hline 1
    \end{pmatrix},
    \begin{pmatrix}
    1\\ \hline 0
    \end{pmatrix},
\end{align}
and a nearest neighbor Ising coupling in the $x$ direction $Z_0Z_{\hat x}$ and $y$ direction $Z_0Z_{\hat y}$ can be represented as
\begin{equation}
\lc
    \begin{pmatrix} 1+x\\ \hline 0\end{pmatrix}, \begin{pmatrix} 1+y\\ \hline 0\end{pmatrix}.
\end{equation}
Here, we use $x_1 =x$, $x_2 =y$, and $x_3=z$.

We remark that addition of the vectors in Eq. \eqref{equ:polynomialform} (modulo 2) corresponds to multiplication in terms of Pauli operators in Eq. \eqref{equ:Pauliform}. For example,
\begin{align}
\lc
    \begin{pmatrix}
    0\\ \hline 1
    \end{pmatrix} +
     \begin{pmatrix}
    1\\ \hline 1
    \end{pmatrix} =
    \begin{pmatrix}
    1\\ \hline 0
    \end{pmatrix},
\end{align}
corresponds to $X_0 Y_0 \propto Z_0$. Furthermore, multiplication by a monomial gives a natural action of translation. For example, translating the operator by $n$ sites in the $j^\text{th}$ spatial direction can be obtained by multiplying the corresponding vector by $x_j^n$. 

Let us call the space of the polynomial representation of all Pauli operators $P$. Mathematically, the polynomials live in a Laurent polynomial ring $R=\mathbb F_2[x_1^{\pm1},...,x_d^{\pm 1}]$, and the action of this ring as translation on $P \cong R^2$ gives it the structure of an $R$-module, called the \textit{Pauli module}.

In general, any lattice can be recast into a cubic lattice with $N$ spins per unit cell, for some $N$. In such a case, we can represent any operator as a length $2N$ vector, with the first $N$ entries denoting the positions of the Pauli $Z$'s, and the next $N$ corresponding to Pauli $X$'s. For example, consider the stabilizers of the 2D toric code
\begin{align}
    \includegraphicsr{1}{2DKW_1form1.pdf},& &  \includegraphicsr{1}{2DKW_Zt.pdf}.
\end{align}
Although the qubits live on edges, we can turn it into a square lattice with two sites per unit cell by identifying each horizontal and vertical edge to the vertex to the immediate left and bottom, respectively. In this case, the stabilizers can be written respectively up to translation as vectors
\begin{align}\lc
\begin{pmatrix}
0\\0\\
\hline
1+y\\1+x
\end{pmatrix}, &&\lc
\begin{pmatrix}
1+\bar x\\1+\bar y\\
\hline
0\\0
\end{pmatrix},
\end{align}
in $P \cong R^4$.

\subsubsection{Hamiltonians and its corresponding generating map}

Next, we would like to represent a general translation-invariant Pauli Hamiltonian in the algebraic formalism. The key point of this procedure is to extract a set of ``generating'' operators which has the same symmetries as the Hamiltonian and can generate the algebra of all the operators in the Hamiltonian.

Let us assume that the Hamiltonian is defined on a cubic lattice with $N$ sites per unit cell expressible by $\tau$ independent terms up to translation:
\begin{align}
    H = \sum_{\bs i \in \mathbb Z^d} w_1H^{(1)}_{\bs i} + \cdots + w_\tau H^{(\tau)}_{\bs i}.
\end{align}
where $H^{(j)}_{\bs i}$ are Pauli operators and $w_j$ are weights. The procedure is given by the following steps
\begin{enumerate}
    \item Find a minimal set of ``generating'' Pauli operators $\sigma^{(1)},...,\sigma^{(T)}$ defined as a subset of $H_{\bs 0}^{j}$ such that no operator in this subset is a product of the other operators in this subset or their respective translations.
    \item Represent the generating operators algebraically as $\bs \sigma^{(1)}, ... ,\ \bs \sigma^{(T)}$ and define the following \textit{generating map}\footnote{In Ref. \onlinecite{Haah2013}, $\bs \sigma$ was called the stabilizer map. Since terms do not necessarily commute, we call it the generating map to avoid confusion.} given by a $2N \times T$ matrix
    \begin{align}
\bs \sigma = \left (\bs \sigma^{(1)}  \ \cdots  \ \bs \sigma^{(T)}\right).
\end{align}
Because we have chosen a minimal set, the columns of $\bs \sigma$ are linearly independent by construction.
    \item Find column vectors $\bs g^{(t)}$ of length $T$ for $t=1,...,\tau$ such that
        \begin{align}
\bs H^{(t)} = \bs \sigma \bs g^{(t)},
\end{align}
where $\bs H^{(t)}$ is the algebraic representation of $H^{(t)}$. We will call $\bs g^{(t)}$ the \textit{generator labels}.
    \item The Hamiltonian has been recasted into a generating map $\bs \sigma$ accompanied by generator labels $\bs g^{(1)},..., \ \bs g^{(\tau)}$ and weights $w_1,...,\ w_\tau$. 
\end{enumerate}

To recover the Hamiltonian from such data, let $G$ denote the set of all the possible generator labels. Then for any element $\bs g \in G$ represented as
\begin{align}
   \bs g= \begin{pmatrix}
    \sum_{\bs i} g^{(1)}_{\bs i} x_1^{i_1} \cdots x_{d}^{i_d}\\
    \vdots\\
    \sum_{\bs i} g^{(T)}_{\bs i} x_1^{i_1} \cdots x_{d}^{i_d}
    \end{pmatrix},
\end{align}
we can define the Pauli operator 
\begin{align}
 \sigma^{(\bs g)}  =   \prod_{\bs i}\left (\sigma^{(1)}_{\bs i}\right)^{g^{(1)}_{\bs i}} \cdots \left (\sigma^{(T)}_{\bs i}\right)^{g^{(T)}_{\bs i}},
\end{align}
where $\sigma^{(t)}_{\bs i}$ is the operator $\sigma^{(t)}$ translated to coordinate $\bs i$. We notice that the algebraic representation of this Pauli operator is exactly $\bs \sigma \bs g$. Therefore, let us define the following Hamiltonian
\begin{align}
    H_{\bs \sigma} = \sum_{\bs i \in \mathbb Z^d}\sum_{\bs g \in G} w_{\bs g} \sigma^{(\bs g)}_{\bs i},
    \label{equ:Hamiltonian}
\end{align}
which we will call the Hamiltonian generated by $\bs \sigma$. This Hamiltonian is more general than the input Hamiltonian as it includes many more terms, but we can see that our original Hamiltonian can be considered a special case of the generated Hamiltonian $H_{\bs \sigma}$ by choosing a particular choice of weights. Namely, by choosing the weights to be $w_{\bs g} = w_t$ for $\bs g$ such that $\bs H^{(t)} =\bs \sigma \bs g$ for some $t=1,...,T$, and zero otherwise.

The action of translation by monomial multiplication on $G$ also makes $G \cong R^T$ an $R$-module, called the \textit{generator-label module}, and the generating map is now well defined as a map $\bs \sigma: G \rightarrow P$.

One might notice that the generated Hamiltonian \eqref{equ:Hamiltonian} for arbitrary nonzero weights $w_{\bs g}$ correspond to the same generating map $\bs \sigma$ with different sets of generator labels $\bs g^{(t)}$. Therefore, it is convenient to study the properties of the generated Hamiltonian (and later, the dualities between them), of which our input Hamiltonian is a special case.

\subsubsection{Commutation relations and the excitation map}
Next, we study commutation relations between a Pauli operator and a generating operator $\sigma^{(t)}$ in the algebraic formalism.

Commutation relations of operators $A$ and $B$ -- algebraically represented as $\bs A$ and $\bs B$ -- at different positions are efficiently calculated via a symplectic inner product 
\begin{equation}
    \inner{\bs A,\bs B} = \bs A^\dagger \bs \lambda_N \bs B
\end{equation}
 called the \textit{commutation value}. Here, the $\dagger$ superscript denotes the matrix transpose, followed by the antipode map $x_i \rightarrow \bar x_i \equiv x_i^{-1}$ (i.e., an inversion with respect to the origin), and $\bs \lambda_N$ is the symplectic form\footnote{Working with coefficients in $\mathbb F_2$, the minus sign can be dropped.}
\begin{align}
\bs \lambda_N = \begin{pmatrix} 0_{N\times N} & \mathbbm 1_{N\times N}\\\mathbbm{1}_{N\times N} & 0_{N\times N}\end{pmatrix}.
\end{align}

The nonzero terms in the commutation value $\inner{\bs A,\bs B}$ denotes the translations of the operator $A$ which anticommute with the operator $B$ placed at the origin. To see this, the coefficient of the monomial $1$ counts the number of Pauli $Z$'s from $A$ that appears in the same position as a Pauli $X$ from $B$ and vice versa (modulo 2). Furthermore, the coefficient of a monomial $m$ of $\inner{\bs A,\bs B}$ is equal to the coefficient of the monomial $1$ of $\inner{m\bs A,\bs B}$, which counts the number of anticommutations of $A$ translated according to $m$, and $B$.

Based on this fact, we can now determine the positions where each operator in the Hamiltonian anticommutes with a given Pauli operator. For this purpose, it is sufficient to determine the commutation value of a Pauli operator $p$ with the generating operators. Therefore, for each $\sigma^{(j)}$ where $j=1,...,T$, we can compute $\inner{\bs \sigma^{(j)},\bs{p}}$ for any $\bs{p} \in P$ and the resulting polynomial for each $j$ determines the translations of the generating operators that anticommute with an operator $p$.

To package this information, it is helpful to introduce the \textit{excitation map} $\bs \epsilon = \bs \sigma^\dagger \bs \lambda_N$. Doing so, we see that $\bs\epsilon(\bs p)$ is a column vector of length $T$ which encodes the commutation values above. Formally, the excitation map is defined as $\bs \epsilon:P \rightarrow E$, where $E \cong R^T$ labels the the positions of anticommutations for each generating operator. With the action of translation, $E$ is also an $R$-module called the \textit{excitation module}. We note that $G$ and $E$ are equivalent as $R$-modules, as they both label the generating operators, but they are named differently for clarity.

To compute the commutation relation between generating operators, we define the \textit{commutation matrix} as $\inner{\bs \sigma, \bs \sigma}$. This is a $T\times T$ matrix whose entries denote the commutation value between generating operators, and thus encodes the commutation relations of all operators in the Hamiltonian.

\subsubsection{Symmetries and identity generators}

Having defined the modules and maps between them, let us summarize them with the following sequence
\begin{equation}
\begin{tikzcd}
 G  \arrow[r,"\bs \sigma" ] & P \arrow[r,"\bs \epsilon"] & E
\end{tikzcd}.
\label{equ:sequence1}
\end{equation}
We now see that the symmetries and identity generators of the Hamiltonian \eqref{equ:Hamiltonian} are conveniently encoded kernel of such maps, namely
\begin{enumerate}
\item The kernel of $\bs \sigma$ denotes the set of generating operators that product to the identity, which are the \textit{identity generators}.
\item The kernel of $\bs \epsilon$ denotes the set of Pauli operators which commute with all the generating operators and hence the Hamiltonian. They therefore denote the \textit{symmetries}.
\end{enumerate}

We remark that for stabilizer Hamiltonians, all terms in the Hamiltonian commute, meaning that $\inner{\bs \sigma, \bs \sigma} =0$. This can be recast in to the statement that Eq. \eqref{equ:sequence1} is a chain complex. That is, $\bs \epsilon \circ \bs \sigma =0$. Physically, this means that all terms in the Hamiltonian obtained by the generating map $\bs \sigma$ commute with one another, and are thus annihilated by the excitation map $\bs \epsilon$.

When the terms in the Hamiltonian do not commute, the sequence \eqref{equ:sequence1} is not a chain complex. Nevertheless, the algebra of all the operators in the Pauli Hamiltonian is encoded in $\inner{\bs \sigma, \bs \sigma}$, which is the map $\bs \epsilon \circ \bs \sigma$. Intuitively, this map inputs a generator label and outputs all the other generator labels and their translations which anticommute with this input.

\subsubsection{Basis transformations}
Operators can be represented differently by changing the basis of the wavefunction. This is implemented in the Heisenberg picture as a unitary operation $U$ which maps between Pauli operators. Let $\bs U$ be the algebraic representation of such a unitary, which is an isomorphism of the Pauli module $P$. Since it must preserve the commutation relations of all Pauli operators represented in $P$, this requires $\inner{\bs U\bs A,\bs U\bs B} = \inner{\bs A,\bs B}$,  which implies that $\bs  U^\dagger \bs \lambda_N \bs U = \bs \lambda_N$. Hence, $\bs U$ is a symplectic transformation. When there is no translation involved, the unitaries that realize such transformation between stabilizer Hamiltonians are Clifford unitaries.  Let us give two examples:
\begin{enumerate}
\item For a lattice with one site per unit cell, consider
\begin{align}
    \bs U = \begin{pmatrix} 1&1\\0&1\end{pmatrix}.
\end{align}
In terms of Pauli matrices, it sends $Z_{\bs i} \rightarrow Z_{\bs i}$ and $X_{\bs i} \rightarrow Z_{\bs i} X_{\bs i} \propto Y_{\bs i}$. Therefore this symplectic matrix corresponds to an $S$ gate acting on all sites, where $S = \text{diag}(1,i)$.
\item For a lattice with two sites per unit cell, consider
\begin{align}
    \bs U = \begin{pmatrix} 
    1&0&0&1\\
    0&1&1&0\\
    0&0&1&0\\
    0&0&0&1
    \end{pmatrix}.
\end{align}
Reading column by column, the Pauli matrices transform as
\begin{align}
 (Z\otimes I)_{\bs i} &\rightarrow (Z\otimes I)_{\bs i}, & (I\otimes Z)_{\bs i} &\rightarrow (I\otimes Z)_{\bs i}, \nonumber\\
    (X\otimes I)_{\bs i} &\rightarrow (X\otimes Z)_{\bs i}, & (I\otimes X)_{\bs i} &\rightarrow (Z\otimes X)_{\bs i}.
\end{align}
Therefore, this symplectic matrix acts as a controlled-$Z$ gate between the two sites in each unit cell.
\end{enumerate}
In this paper, we will only need to use Clifford phase gates, which are generated by $S$ and controlled-$Z$. In the algebraic representation, these turn out to be matrices of the form
\begin{align}
\bs U = \begin{pmatrix} 
    \mathbbm 1_{N \times N} & \bs A_{N \times N}\\
    0_{N \times N} & \mathbbm 1_{N \times N}
    \end{pmatrix}.
\end{align}
For a Hermitian matrix $\bs A = \bs A^\dagger$. In fact, one can verify that the above matrix is only symplectic when $\bs A$ is Hermitian\footnote{The phase gate above is most intuitively depicted by edges of a translation-invariant graph that connect vertices at the origin to the locations determined by $\bs A$. The unitary consists of $S$ gates where edges connect the same vertex, and controlled-$Z$ gates on all other edges. Translation invariance requires that any edge leaving a vertex must also have a corresponding edge leaving in the opposite direction. This is the condition that $\bs A$ is Hermitian.}.

\subsubsection{Example: 1D transverse-field Ising model}

Let us demonstrate with an example of operators in a 1D transverse-field Ising model.
\begin{equation}
    H = \sum_i Z_{i}Z_{i+1} + h X_{i}
\end{equation}
The generating Pauli operators can be chosen to be the Ising coupling $Z_0Z_1$ and the transverse field $X_0$. In the polynomial representation, they are
\begin{align}
 \bs Z &=\lc \begin{pmatrix} 1+x\\ \hline 0\end{pmatrix}, & \bs X &= \lc\begin{pmatrix} 0\\ \hline 1\end{pmatrix}.
 \label{equ:1DIsing}
\end{align}
Here (and throughout the paper), we have reserved the symbol $\bs Z$ for the Ising terms, (rather than for Pauli $Z$). Since the two operators are independent, the generating map corresponding to this Hamiltonian is
\begin{align}
\lc
    \bs \sigma = (\bs Z \ \bs X) = \begin{pmatrix} 1+x & 0\\ \hline 0 &1\end{pmatrix}.
\end{align}
and the two terms in the Hamiltonian correspond to generator labels $\lc \bs g = \begin{pmatrix} 1\\ 0\end{pmatrix}$ and $\bs g = \lc \begin{pmatrix} 0\\  1\end{pmatrix}$ with weights $w_1=1$ and $w_2=h$.

Next, we compute the commutation value of $\bs Z$ and $\bs X$
\begin{align}
    \inner{\bs Z,\bs X}= \bs Z ^\dagger \bs \lambda_1 \bs X= \begin{pmatrix}[c!{\color{\linecolor}\vrule} c] 1+\bar x & 0
    \end{pmatrix} \begin{pmatrix}0&1\\1&0
    \end{pmatrix}\lc \begin{pmatrix} 0\\ \hline 1
    \end{pmatrix} = 1 + \bar x,
\end{align}
This means that each transverse field $X_i$ anticommutes with the Ising coupling $Z_iZ_{i+1}$ and the one shifted one site to the left ($Z_{i-1}Z_i$).

The corresponding excitation map is
\begin{equation}
    \bs \epsilon = \bs \sigma^\dagger\bs\lambda_1=   \begin{pmatrix}[c!{\color{\linecolor}\vrule} c] 0 &1 +\bar x\\ 1&0\end{pmatrix}.
\end{equation}
This can be used to determine the Pauli operators that anticommute with the generating operators. For example, a single Pauli $Z$ at the origin, represented as $\lc \begin{pmatrix} 1 \\ \hline 0 \end{pmatrix}$, maps to
\begin{equation}
  \begin{pmatrix}[c!{\color{\linecolor}\vrule} c] 0 &1 +\bar x\\ 1&0\end{pmatrix} \lc \begin{pmatrix} 1 \\ \hline 0 \end{pmatrix}= \begin{pmatrix}
    0\\ 1
  \end{pmatrix},
\end{equation}
meaning it anticommutes with the second generating operator located at the origin, which is the transverse field $X_0$. 

Finally, let us look at the symmetries and identity generators. The symmetry is given by the kernel of the above excitation map. Here, we will slightly abuse notation and represent the kernel of a map via its generators in the algebraic formalism (that is, the kernel is written as the columns which span it). In this case, there is only one generator\footnote{This statement is actually not mathematically precise, since an infinite sum is not contained in $\mathbb F_2[x^{\pm 1}]^2$. However, all subsequent statements can be made precise by instead considering a periodic system with module $R=(\mathbb F_2[x_1,..., x_d]/ ( x_1^L-1,...,x_d^L-1 ) )^{2K}$, and replacing the infinite sum with bounds from $0$ to $L-1$, for some system size $L$ much larger than the range of all local interactions}:
\begin{equation}
\lc
    \ker \bs \epsilon =   \begin{pmatrix} 0 \\ \hline \sum_i x^i\end{pmatrix} = \bs X\sum x^i.
\end{equation}
As Pauli matrices, this is $\prod_i X_i$, the $\mathbb Z_2$ global spin-flip symmetry of the Ising model.

The identity generators can be seen from the fact that all the Ising couplings along the 1D chain product to the identity.
\begin{align}
    \prod_i Z_i Z_{i+1}=1
\end{align}
Algebraically, this corresponds to
\begin{equation}
\bs \sigma \begin{pmatrix}\sum_i x^i\\0 \end{pmatrix} = \bs Z \sum_i x^i =0.
\end{equation}
In fact, this is the only identity generator and therefore we have that
\begin{equation}
\ker \bs \sigma  =\begin{pmatrix}\sum_i x^i\\0.
\end{pmatrix}
\label{equ:identityrelation1D}
\end{equation}

As an application, we can expand the tunable parameters to this Hamiltonian. For example, we can consider
\begin{equation}
    H' = \sum_i Z_{i}Z_{i+1} +h X_{i} - h' Y_iY_{i+1} +h'' Z_{i-1}X_iZ_{i+1},
    \label{equ:long1DHam}
\end{equation}
which has a much richer phase diagram and phase transitions\cite{VerresenMoessnerPollmann2017,VerresenThorngrenJonesPollmann2019}. Since, all terms in the Hamiltonian are appropriate products of the Ising term and the transverse field, it has the same generating map, which implies that the Hamiltonian also has the same symmetries and identity generators. Importantly, the KW and JW dualities considered later in this paper will allow us to not only dualize the transverse-field Ising model but also more complicated Hamiltonians such as the above.

\subsection{KW Duality between Generalized transverse-field Ising Models}\label{transverseIsing}
Let us now define a generalized transverse-field Ising model (which we will refer to as ``Ising model'' for brevity) within the algebraic formalism.  Consider two sets of operators, which we will call \textit{transverse fields} and \textit{Ising interactions}, denoted algebraically by $\bs X$ and $\bs Z$, respectively. We denote $K$ the number of sites per unit cell (and hence the number of independent transverse fields) and $N$ the number of independent types of Ising interactions up to multiplication of operators and translation. Without loss of generality, we choose a basis in which the transverse fields point in the $x$ direction. Hence, we can algebraically represent the transverse fields as $K$ single Pauli $X$'s
\begin{align}
   \bs X &=\lc\begin{pmatrix}0_{K\times K} \\ \hline \mathbbm 1_{K\times K}\end{pmatrix}.
\end{align}
On the other hand, the $N$ Ising terms can be arbitrary interactions that all mutually commute but do not commute with the transverse fields. Thus, $\bs Z$ is a $2K\times N$ matrix. Together with $\bs X$, they satisfy
\begin{align}
   \inner{\bs Z,\bs Z}& = \inner{\bs X, \bs X}=0,   & \inner{\bs Z,\bs X} \ne 0.
\end{align}
The Hamiltonian for such an Ising model can be represented by the generating map
\begin{align}
    \bs \sigma = (\bs Z \ \bs X).
\end{align}
That is, the first $N$ generating operators are the Ising terms, and the next $K$ generating operators are the transverse fields. Because they do not commute, it is evident that the commutation matrix
\begin{equation}
    \inner{\bs \sigma, \bs \sigma} =\begin{pmatrix} 0_{N\times N} &\inner{\bs Z,\bs X}\\ \inner{\bs X,\bs Z}& 0_{K \times K}\end{pmatrix}
    \label{equ:sigmacommutation}
\end{equation}
is nonzero.

Given an Ising model represented by $\bs \sigma$, one could ask whether there exists another set of operators, which have the same commutation relations as those of the given Ising model.  That is, whether there is a \textit{dual} generating map $\tbs \sigma: G \rightarrow \tilde P$, such that
$\inner{\bs \sigma, \bs \sigma} =\inner{\tbs \sigma, \tbs \sigma} $. Defining $\tbs \epsilon = \tbs \sigma^\dagger \bs \lambda_N$ and recalling that $\inner{\bs \sigma, \bs \sigma} = \bs \epsilon \circ \bs \sigma$, this is equivalent to finding a $\tbs \sigma$ such that the diagram

\begin{equation}
\begin{tikzcd}
 G  \arrow[rd, "\tbs \sigma"]\arrow[r,"\bs \sigma" ] & P \arrow[r,"\bs \epsilon"] & E\\
 & \tilde P \arrow[ru, "\tbs \epsilon"]& 
\end{tikzcd}
\label{equ:sequence2}
\end{equation}
commutes.

We will restrict our attention to dualities between generalized Ising models. Here, we define the KW dual to be a certain specific choice of the dual generating map $\tbs \sigma$, which itself represents another generalized Ising model. The dual map is given by $\tbs \sigma=(\tbs X \ \tbs Z) $, where
\begin{align}
   \tbs X &=\lc\begin{pmatrix}0_{N\times N} \\ \hline \mathbbm 1_{N\times N}\end{pmatrix}, & \tbs Z &=\lc\begin{pmatrix} \inner{\boldsymbol Z,\boldsymbol X}\\\hline 0_{N\times K}\end{pmatrix}.
   \label{equ:KWdualoperators}
\end{align}
This formally describes the procedure of obtaining the KW dual operators given in Sec. \ref{review}. Importantly, the role of the transverse fields and Ising terms in the dual model are swapped: There are now $N$ sites per unit cell (hence $N$ transverse field terms) and $K$ Ising terms. Furthermore, it follows that $P$ and $\tilde P$ are different modules (meaning the two Ising models can live in different Hilbert spaces) when $N \ne K$.

Let us verify that the choice above is a valid dual Ising model. That is, the commutation relations are preserved.
\begin{prop} The map $\tbs \sigma=(\tbs X \ \tbs Z)$ is a valid KW dual.
\begin{proof}
Since $\inner{\tbs Z,\tbs Z}= \inner{\tbs X,\tbs X}=0$, and $\inner{\tbs Z,\tbs X} = ( \inner{\bs Z,\bs X}^\dagger {\color{red}|} 0_{K\times N} )\bs \lambda_N \lc\begin{pmatrix} 0_{N\times N}\\ \hline \mathbbm 1_{N\times N}\end{pmatrix} = \inner{\bs X,\bs Z}$. Therefore,
\begin{align}
\lc
    \inner{\tbs \sigma, \tbs \sigma} &=\begin{pmatrix} \inner{\tbs X,\tbs X}  &\inner{\tbs X,\tbs Z}\\ \inner{\tbs Z,\tbs X}& \inner{\tbs Z,\tbs Z} \end{pmatrix} \nonumber \\
    &=\begin{pmatrix} 0_{N \times N}  &\inner{\bs Z,\bs X}\\ \inner{\bs X,\bs Z}& 0_{K \times K} \end{pmatrix} =  \inner{\bs \sigma, \bs \sigma}
\end{align}
as desired. 
\end{proof}
\end{prop}
\begin{table*}[t!]

\caption{Mapping of important operators under KW duality in the algebraic formalism. Here, the set of important generator labels form a submodule $\frak g \subset G$, which correspond to the set of operators $\bs \sigma \frak g$ in the Ising model and operators $\tbs \sigma \frak g$ in the dual Ising model.  In general, the operator $\bs \sigma \bs g$ is dual to $\tbs \sigma \bs g$ for any $\bs g \in G$.}
\begin{tabular}{|c|c|r l|c| r l|c|}
\hline
 \multirow{2}{*}{Generator labels ($\frak g$)}& \multirow{2}{*}{Note} & \multicolumn{3}{c|}{Ising model ($\bs \sigma = (\bs Z \ \bs X)$)} &  \multicolumn{3}{c|}{Dual Ising model ($\tbs \sigma = (\tbs X \ \tbs Z)$)}\\
 \cline{3-8}
 &&\multicolumn{2}{c|}{Operators ($\bs \sigma \frak g$)} & Note & \multicolumn{2}{c|}{Operators ($\tbs \sigma \frak g$)}& Note\\
 \hline
 $\begin{pmatrix} \mathbbm 1_{N\times N } \\ 0_{K\times N} \end{pmatrix}$&First $N$ generators&$\bs Z$&   &Ising terms &  $\tbs X$ &$=\lc\begin{pmatrix}0_{N\times N} \\ \hline \mathbbm 1_{N\times N}\end{pmatrix}$ & Dual transverse field \\
\hline
 $\begin{pmatrix}  0_{N\times K } \\ \mathbbm 1_{K\times K} \end{pmatrix}$&Next $K$ generators&$\bs X$&$= \lc\begin{pmatrix}0_{K\times K} \\ \hline \mathbbm 1_{K\times K}\end{pmatrix}$   &Transverse field &  $\tbs Z$& $=\lc\begin{pmatrix} \inner{\boldsymbol Z,\boldsymbol X}\\\hline 0_{N\times K}\end{pmatrix}$ & Dual Ising terms \\
 \hline
$\ker \bs \sigma$&Identity generators & 0& & Identity  & $\ker \tbs \epsilon$&$= \tbs \sigma \ker \bs \sigma$ & Dual symmetry\\
\hline
$\ker \tbs \sigma$&Dual identity generators & $\ker \bs \epsilon$&$= \bs \sigma \ker \tbs \sigma$ & Symmetry & 0&  & Identity \\
\hline
\multicolumn{8}{c}{Condition: $\inner{\bs Z,\bs Z}=0$}
\end{tabular}
\label{tab:KWsummary}
\end{table*}

We can now see how to map any operator in the generated Hamiltonian \eqref{equ:Hamiltonian} to its dual generated Hamiltonian. Since a valid term in the generated Hamiltonian lives in $\im \bs \sigma$, there exists a generator label $\bs g \in G$ which represents this term, and its corresponding dual can be given by $\tbs \sigma \bs g$. Said differently, for each $\bs g \in G$, $\bs \sigma \bs g \in P$ is dual to $\tbs \sigma \bs g \in \tilde P$. It is important to again emphasize that not all Pauli operators are mapped under the duality. Only those that are generated by operators appearing in the Hamiltonian are dualizable. In the algebraic language, the isomorphism is not between $P$ and $\tilde P$, but between $\im \bs \sigma$ and $\im \tbs \sigma$, which are submodules of $P$ and $\tilde P$, respectively.

We conclude by stating the correspondence between symmetries and identity generators. Mathematically, the identity generators ($\ker \bs \sigma$) can be dualized by mapping through the dual generating map $\tbs \sigma$. Therefore, the correspondence amounts to the following statements:
\begin{prop}  $\ker \bs \epsilon= \bs \sigma \ker \tbs \sigma$ \label{prop:KWconstraint1}
\end{prop}
\begin{prop} $\ker \tbs \epsilon=\tbs \sigma \ker \bs \sigma$ \label{prop:KWconstraint2}
\end{prop}
We provide proofs of the above in Appendix \ref{app:KWJWconstraints}.\\
The relevant operators under the KW duality are summarized in Table \ref{tab:KWsummary}. For completeness, we also summarize the explicit expression of the generating and excitation maps.
\begin{align}
   \bs  \sigma &= (\bs Z \ \bs X), &\tbs  \sigma &= (\tbs X \ \tbs Z) = \lc\begin{pmatrix}  0_{N \times N}& \inner{\boldsymbol Z,\boldsymbol X} \\\hline \mathbbm 1_{N\times N} & 0_{N\times K}\end{pmatrix}, \nonumber\\
   \bs \epsilon &= \bs \sigma^\dagger \bs \lambda_{K}, & \tbs \epsilon &=\tbs \sigma^\dagger \bs \lambda_{N} = \begin{pmatrix}[c!{\color{\linecolor}\vrule} c] \mathbbm 1_{N\times N} & 0_{N\times N}\\
  0_{K\times N} &\inner{\boldsymbol X,\boldsymbol Z} \end{pmatrix}.
\end{align}

\subsubsection{Example: KW self-duality in the 1D transverse-field Ising model}

Let us compute the dual generating operators of the 1D Ising model. Inserting Eq. \eqref{equ:1DIsing} into Eq. \eqref{equ:KWdualoperators}, the dual operators are
\begin{align}
 \tbs X &= \lc\begin{pmatrix} 0\\ \hline 1\end{pmatrix}, & \tbs Z &= \lc\begin{pmatrix} 1+\bar x\\ \hline 0\end{pmatrix}.
 \label{equ:1DIsingdual}
\end{align}
which are the same set of operators up to a translation of $\tbs Z$ by $x$. Therefore, the 1D Ising model is self-dual under the KW duality.

We can also show the correspondence between symmetries and identity generators.
From Prop. \ref{prop:KWconstraint2}, we can calculate the dual symmetry as
\begin{equation}
\lc
\ker \tbs \epsilon = \tbs \sigma \ker \bs \sigma =   \begin{pmatrix}0 & 1+\bar x \\ 
\hline 1 & 0 \end{pmatrix}\begin{pmatrix}\sum_i x^i\\0 \end{pmatrix} = \begin{pmatrix} 0\\ \hline \sum_i x^i\end{pmatrix},
\end{equation}
which is the global $\mathbb Z_2$ symmetry as expected.

Finally, let us demonstrate how to dualize the Hamiltonian \eqref{equ:long1DHam}. Using the same generating operators, we see that the corresponding generator labels for the four terms are, respectively,
\begin{align}
    \bs g = \begin{pmatrix} 1\\0 \end{pmatrix}, \ \begin{pmatrix} 0\\1 \end{pmatrix}, \ \begin{pmatrix} 1\\1+x \end{pmatrix}, \ \begin{pmatrix} 1+\bar x\\1 \end{pmatrix}.
\end{align}
Therefore, using $\tbs \sigma = (\tbs X \ \tbs Z)$, the dual operators are algebraically
\begin{align}
  \lc  \tbs \sigma \bs g =\begin{pmatrix} 0\\\hline 1 \end{pmatrix}, \ \begin{pmatrix} 1+\bar x\\\hline 0 \end{pmatrix}, \ \begin{pmatrix} x+\bar x\\\hline  1 \end{pmatrix}, \ \begin{pmatrix} 1+\bar x\\\hline 1+\bar x \end{pmatrix}.
\end{align}
Translating back to Pauli operators, the dual Hamiltonian is given by
\begin{align}
    \tilde H= \sum_i  X_{i} + h  Z_{i-1}Z_{i} + h' Z_{i-1}X_iZ_{i+1} -h'' Y_{i-1}Y_{i}.
\end{align}

\subsection{Example: transverse-field Ising model on a square lattice}\label{KW2D}
For completeness, we demonstrate how the KW duality on a square lattice given in Sec. \ref{KWreview} is done in the algebraic formalism. Given the Hamiltonian \eqref{equ:2DTFIM}, its corresponding generating and excitation maps are
\begin{align}
   \ \bs \sigma & = \lc \begin{pmatrix} 1+  x & 1+ y & 0\\\hline 0 &0 & 1 \end{pmatrix}, & \bs \epsilon & = \begin{pmatrix}[c!{\color{\linecolor}\vrule} c] 0 & 1+\bar x\\
    0 & 1+\bar y\\
    1& 0 \end{pmatrix}.
\end{align}
The global $\mathbb Z_2$ symmetry $\prod_{i,j} X_{i,j}$ is given by
\begin{align}
\lc
\ker \bs \epsilon = \begin{pmatrix} 0\\ \hline \sum_{ij}x^iy^j\end{pmatrix},
\end{align}
and the identity generators are
\begin{align}
    \ker \bs \sigma = \begin{pmatrix}
1+y & \sum_i x^i &0 \\
1+x & 0 &\sum_i y^i \\
0 & 0 & 0
\end{pmatrix},
\end{align}
where the first column corresponds to the product of Ising terms around a plaquette, and the second and third are a product around the horizontal and vertical cycles of the torus.

Using the KW duality \eqref{equ:KWdualoperators}, the dual operators are
\begin{align}
   \tbs X &=\lc\begin{pmatrix}    
   0 & 0 \\
    0 & 0 \\
    \hline  
    1 &0\\
   0&1 \end{pmatrix}, &\tbs Z &=\lc\begin{pmatrix}  1+ \bar x\\
    1+ \bar y\\
    \hline
    0\\
    0\
    \end{pmatrix}.
\end{align}
which are the terms in the dual Hamiltonian \eqref{equ:2DKWdualHam}. The dual generating and excitation maps are
\begin{align}
   \tbs \sigma &=\lc \begin{pmatrix}    
   0 & 0 & 1+ \bar x\\
    0 & 0 &1+\bar y\\
    \hline  
    1 &0 &0\\
   0&1 &0 \end{pmatrix}, &
   \tbs \epsilon &=\begin{pmatrix}  [cc!{\color{\linecolor}\vrule} cc]
     1 &0  &0 & 0 \\
        0&1&0 & 0 \\
  0&0 &  1+x & 1+  y
    \end{pmatrix}.
\end{align}
Using Prop. \ref{prop:KWconstraint2}, we can obtain the dual symmetry
\begin{align}
\ker \tbs \epsilon &=\tbs \sigma \ker \bs \sigma =\lc\begin{pmatrix}
0 & 0 &0 \\
0 & 0 &0 \\
\hline
1+ y & \sum_i x^i &0\\
1+ x &0  & \sum_j y^j\\
\end{pmatrix},
\end{align}
which are precisely the 1-form symmetries in Eq. \eqref{equ:1formsym}.

\section{JW Dualities in the Algebraic Formalism}\label{JWduality}
In this section, we will generalize the KW formalism described in order to bosonize a fermionic model into a spin model. We review a similar algebraic formalism for fermions in Sec. \ref{fermionpolynomial}. With this notation, we write down a fermionic model with generic $q$-body interactions for any even $q$ and propose the JW duality to spin systems in Sec. \ref{JW}. The existence and uniqueness up to a basis of this duality is further elaborated in Sec. \ref{JWproofs}. We remark that the new dualities derived in this paper with fracton excitations are those where free-fermion hopping terms ($q=2$) are prohibited.
\subsection{Algebraic Formalism for Fermions}\label{fermionpolynomial}
Hamiltonians for translation-invariant fermions can also be efficiently represented in the algebraic representation\cite{VijayHaahFu2015}. Because the formalism is nearly identical, we will mainly note the main changes that must be made compared to Sec. \ref{polynomialboson}.

Given a complex fermion at each site on a $d$-dimensional cubic lattice, operators can be expressed up to a phase in the algebraic formalism by representing the position of the Majorana operators $\gamma$ and $\gamma'$. Explicitly, a fermion operator $p$ can be decomposed as
\begin{align}
    p \propto \bigotimes_{\bs i} \gamma_{\bs i}^{a_{\bs i}} {\gamma'_{\bs i}}^{b_{\bs i}},
\end{align}
where we can neglect the order of this product by ignoring the associated phase factor. Similarly, we will represent the fermion operator algebraically as the vector
\begin{equation}
\bs{p}=\lc
 \begin{pmatrix}
    \sum_{\bs i}  a_{\bs i}  x_1^{i_1} \cdots x_{d}^{i_d} \\ \hline \sum_{\bs i}  b_{\bs i}  x_1^{i_1} \cdots x_{d}^{i_d}\end{pmatrix}.
\end{equation}
For example, the local fermion parity at the origin $P_0 = -i\gamma_0\gamma_0'$ can be represented as
\begin{equation}
    \bs P = \begin{pmatrix} 1\\ \hline 1\end{pmatrix}.
\end{equation}
In general, when there are $K$ sites per unit cell, we can represent operators as a length $2K$ vector. In such case, the set of $K$ fermion parity operators at the origin are
\begin{equation}
    \bs P =\lc \begin{pmatrix} \mathbbm1_{K\times K}\\ \hline  \mathbbm1_{K\times K}\end{pmatrix}.
\end{equation}

Formally, these vectors live in a module $M$, called the \textit{Majorana module}.

Because the algebra of fermionic operators is $\mathbb Z_2$ graded, while bosonic operators are not, we can only find a mapping between parity even operators to bosonic operators. Therefore, we will from now on assume that all fermionic operators are even.

A general translation-invariant fermionic Hamiltonian with $N$ sites per unit cell can be converted into the algebraic formalism by constructing the generating map $\bs \sigma_F:G \rightarrow M$ whose columns can generate all the terms in the Hamiltonian, and the generator labels are similarly labeled in the generator label module $G$.

Commutation relations between fermionic operators are similarly captured via the commutation value. Given two even operators represented in the algebraic representation as $\bs A$ and $\bs B$, their commutation value is given by the inner product
\begin{equation}
    \inner{\bs A,\bs B}_F = \bs A^\dagger \bs B.
\end{equation}
Here, the subscript $F$ stands for ``fermion'' and denotes the orthogonal as opposed to symplectic inner product used in the bosonic case. Descriptively, this inner product counts the number of overlapping $\gamma$'s and $\gamma'$'s between $A$ and $B$. This is the number of commutations that gets subtracted from the total number of anticommutations when $A$ is commuted through $B$. Like the Pauli case, this commutation value is precisely the translations of the operator $A$ which anticommute with the operator $B$. A rigorous proof can be found in the supplementary material of Ref. \onlinecite{VijayHaahFu2015}.

We can now construct the excitation map, which is given by the adjoint of $\bs \sigma_F$ under the orthogonal inner product. Defining $\bs \epsilon_F = \bs \sigma_F^\dagger$, the excitation map is a map from the Majorana module to the excitation module. $\bs \epsilon_F: M \rightarrow E$. The modules and maps between them can be summarized by the following sequence

\begin{equation}
\begin{tikzcd}
 G  \arrow[r,"\bs \sigma_F" ] & M \arrow[r,"\bs \epsilon_F"] & E
\end{tikzcd}
\label{equ:sequenceF}
\end{equation}
The identity generators and symmetries of the fermionic Hamiltonians are again the kernels of $\bs \sigma_F$ and $\bs \epsilon_F$, respectively.

To summarize, the only main changes from bosons to fermions are a redefinition of symbols and the type of inner product.

\subsubsection{Example: 1D p-wave superconductor}

Let us demonstrate how to convert the Hamiltonian of the 1D toy model for a $p$-wave superconductor\cite{Kitaev2001} into the generating map. The Hamiltonian is given by
\begin{equation}
    H=\sum_{i}( -tc^\dagger_ic_{i+1} + \Delta c_ic_{i+1}+h.c.) -\mu \left (c^\dagger_ic_i-\frac{1}{2}\right )
\end{equation}
In terms of Majorana fermions,
\begin{equation}
    H=\frac{i}{2}\sum_{i}(\Delta+t) \gamma'_{i-1} \gamma_{i} + (\Delta-t) \gamma_{i-1} \gamma_{i}' -\mu  \gamma_i \gamma_i'.
    \label{equ:Majoranachain}
\end{equation}
The generators can be chosen to be the onsite fermion parity $P=-i\gamma_0 \gamma_0'$ and the nearest-neighbor Majorana hopping term $S=i\gamma'_{-1} \gamma_{0}$. We do not need to include the second type of hopping $\gamma_{-1} \gamma_{0}'$ because it can be generated by a product of the two former operators. Algebraically, the two generating operators are
\begin{align}
        \bs S &= \lc \begin{pmatrix} 1\\ \hline \bar x \end{pmatrix}, &\bs P &= \lc \begin{pmatrix} 1\\ \hline 1 \end{pmatrix}.
\end{align}
Therefore, the generating map is given by
\begin{align}
\lc
            \bs \sigma_F = (\bs S \ \bs P) = \begin{pmatrix} 1 & 1\\ \hline \bar x &1 \end{pmatrix},
\end{align}
and the nonzero weights in the Hamiltonian \eqref{equ:Majoranachain}, respectively, correspond to generator labels
\begin{align}
\lc \bs g = \begin{pmatrix} 1\\  0\end{pmatrix}, \begin{pmatrix} 0\\  1\end{pmatrix}, \begin{pmatrix} 1\\  1+\bar x\end{pmatrix}.
\end{align}

The commutation value of $\bs S$ and $\bs P$ is
\begin{align}
    \inner{\bs S,\bs P}_F = \begin{pmatrix}[c!{\color{\linecolor}\vrule} c] 1 & x
    \end{pmatrix} \lc\begin{pmatrix} 1\\ \hline 1
    \end{pmatrix} = 1 +  x,
\end{align}
which means that each fermion parity operator $(P_i =-i\gamma_i\gamma_i')$ anticommutes with the hopping operator $i\gamma_{i-1}\gamma'_{i}$, and the one shifted one site to the right ($i\gamma_{i}\gamma'_{i+1}$).

The corresponding excitation map is
\begin{equation}
    \bs \epsilon = \bs \sigma_F^\dagger=   \begin{pmatrix}[c!{\color{\linecolor}\vrule} c] 1 &x\\ 1&1\end{pmatrix}.
\end{equation}
Therefore, the symmetry is given by
\begin{equation}
\lc
    \ker \bs \epsilon =   \begin{pmatrix} \sum_i x^i \\ \hline \sum_i x^i\end{pmatrix} = \bs P\sum_i x^i.
\end{equation}
In terms of operators (up to a phase), this is $\prod_i P_i$, which is the global fermion parity.

The identity generators can be seen from the fact that up to a phase, the product of all hopping operators and local fermion parity operators is the identity.
\begin{align}
    \prod_i (i\gamma_{i-1}'\gamma_i)(-i \gamma_{i}\gamma_{i}') \propto 1
\end{align}
Algebraically, this corresponds to
\begin{equation}
\bs \sigma \begin{pmatrix}\sum_i x^i\\\sum_i x^i \end{pmatrix} = (\bs S+\bs P) \sum_i x^i =0.
\end{equation}
Therefore,
\begin{equation}
\ker \bs \sigma  =\begin{pmatrix}\sum_i x^i\\\sum_i x^i \end{pmatrix}.
\end{equation}

\begin{table*}[t!]
\caption{Mapping of important operators under JW duality in the algebraic formalism. Here, the set of important generator labels form a submodule $\frak g \subset G$, which correspond to the set of operators $\bs \sigma_F \frak g$ in the fermion model and operators $\tbs \sigma \frak g$ in the dual Pauli model. Note that $\tbs X$ can be considered as a dual transverse field only if $\bs T=0$.  In general, the operator $\bs \sigma_F \bs g$ is dual to $\tbs \sigma \bs g$ for any $\bs g \in G$.}
\begin{tabular}{|c|c|r l|c| r l|c|}
\hline
 \multirow{2}{*}{Generator labels ($\frak g$)}& \multirow{2}{*}{Note} & \multicolumn{3}{c|}{Fermion model ($\bs \sigma_F = (\bs S \ \bs P)$)} &  \multicolumn{3}{c|}{Dual Pauli model ($\tbs \sigma = (\tbs X \ \tbs Z)$)}\\
 \cline{3-8}
 &&\multicolumn{2}{c|}{Operators ($\bs \sigma_F \frak g$)} & Note & \multicolumn{2}{c|}{Operators ($\tbs \sigma \frak g$)}& Note\\
 \hline
 $\begin{pmatrix} \mathbbm 1_{N\times N } \\ 0_{K\times N} \end{pmatrix}$&First $N$ generators&$\bs S$&   &Interaction terms &  $\tbs X$ &$=\lc\begin{pmatrix}\bs T_{N\times N} \\ \hline \mathbbm 1_{N\times N}\end{pmatrix}$ & \\
\hline
 $\begin{pmatrix}  0_{N\times K } \\ \mathbbm 1_{K\times K} \end{pmatrix}$&Next $K$ generators&$\bs P$&$= \lc\begin{pmatrix}\mathbbm 1_{K\times K} \\ \hline \mathbbm 1_{K\times K}\end{pmatrix}$   &Onsite Fermion parity &  $\tbs Z$& $=\lc\begin{pmatrix} \inner{\boldsymbol S,\boldsymbol P}_F\\\hline 0_{N\times K}\end{pmatrix}$ & Dual Ising terms \\
 \hline
$\ker \bs \sigma_F$&Identity generators & 0& & Identity  & $\ker \tbs \epsilon$&$= \tbs \sigma \ker \bs \sigma_F$ & Dual symmetry\\
\hline
$\ker \tbs \sigma$&Dual identity generators & $\ker \bs \epsilon_F$&$= \bs \sigma_F \ker \tbs \sigma$ & Symmetry & 0&  & Identity \\
\hline
\multicolumn{8}{c}{Condition: $ \inner{\bs S, \bs S}_F = \bs T+ \bs T^\dagger$}
\end{tabular}
\label{tab:JWsummary}
\end{table*}

\subsection{Jordan-Wigner Duality}\label{JW}
In the same way that the local fermion parity acts as the transverse field in the Ising model, a fermionic Hamiltonian can have interaction terms that play an analogous role to the Ising interactions. These interaction terms can be thought of as many-body (or correlated) ``hopping'' of fermions.

Let us define an interaction term to be any even $q$-body interaction in the Hamiltonian which does not commute with local fermion parity operators. A fermion system with only $q=2$ interactions is a free-fermion system, but in general we allow $q$ to be any even number (which we will from now on assume). Therefore, each interaction term can be represented as a column of a matrix $\bs S$ which satisfies $\inner{\bs P,\bs S}_F \ne 0$. The number $q$ is equal to the number of monomials appearing in each column of $\bs S$, and in general can vary between columns. We will now study Hamiltonians generated from generating maps of the form $\bs \sigma_F = (\bs S \ \bs P)$ and the duality to a spin Hamiltonian. 

In the same spirit as the KW duality, the JW duality maps fermionic operators to dual Pauli operators which have the same commutations relations. In the algebraic language, given a (fermionic) generating map $\bs \sigma_F$, we would like to find a dual (Pauli) generating map $\tbs \sigma$ such that $\inner{\bs \sigma_F, \bs \sigma_F}_F =\inner{\tbs \sigma, \tbs \sigma} $. That is, the following diagram commutes

\begin{equation}
\begin{tikzcd}
 G  \arrow[rd, "\tbs \sigma"]\arrow[r,"\bs \sigma_F" ] & M \arrow[r,"\bs \epsilon_F"] & E\\
 & \tilde  P \arrow[ru, "\tbs \epsilon"]& 
\end{tikzcd}.
\label{equ:sequence3}
\end{equation}

However, interaction terms, unlike the Ising interactions in the KW duality, do not generally need to commute\cite{ChenKapustinRadicevic2018,ChenKapustin2019,Chen2019}. That is, one might have $\inner{\bs S,\bs S} \ne 0$. Nevertheless, we will demonstrate that such a local mapping from even fermionic operators to bosonic operators always exists. Let us first write down the formal expression of the dual operators corresponding to the descriptive procedure given in \ref{JWreview}. The dual generating map is given by $\tbs \sigma = (\tbs X \ \tbs Z)$, where
\begin{align}
   \tbs X &=\lc \begin{pmatrix}\bs T_{N\times N} \\ \hline \mathbbm 1_{N\times N}\end{pmatrix} & \tbs Z &=\lc \begin{pmatrix} \inner{\boldsymbol S,\boldsymbol P}_F\\\hline 0_{N\times K}\end{pmatrix}.
   \label{equ:XZforJW}
\end{align}
for some matrix $\bs T$ which satisfies $\bs T+ \bs T^\dagger = \inner{\bs S, \bs S}_F$. We will call $\bs T$ the \textit{transmutation matrix}, which attaches additional Pauli $Z$'s at certain positions to each Pauli $X$ dual to the interaction terms. As the name suggests, the importance of $\bs T$ is to modify the statistics of the excitations of the dual model from bosons to fermions when $\inner{\bs S, \bs S}_F \ne 0$. Let us prove that the given map works.

\begin{prop} $\tbs \sigma= (\tbs X \ \tbs Z)$ as given by Eq. \eqref{equ:XZforJW} where $\bs T$ satisfies $\bs T + \bs T^\dagger = \inner{\bs S, \bs S}_F$ is a valid JW dual.
\begin{proof}
We compute  $\inner{\tbs Z,\tbs Z}=0$, $\inner{\tbs X,\tbs X}=\bs T + \bs T^\dagger$, and $\inner{\tbs Z,\tbs X} = (  \inner{\bs S,\bs P}_F^\dagger{\color{\linecolor}|}0_{K\times N} )\bs \lambda_N \lc \begin{pmatrix} \bs T_{N\times N}\\ \hline \mathbbm 1_{N\times N}\end{pmatrix} = \inner{\bs P,\bs S}_F$. Therefore,
\begin{align}
    \inner{\tbs \sigma, \tbs \sigma} &=\lc\begin{pmatrix} \inner{\tbs X,\tbs X}  &\inner{\tbs X,\tbs Z}\\\hline \inner{\tbs Z,\tbs X}& \inner{\tbs Z,\tbs Z} \end{pmatrix} \nonumber \\
    &=\lc\begin{pmatrix} \bs T + \bs T^\dagger  &\inner{\bs S,\bs P}_F\\\hline \inner{\bs P,\bs S}_F& 0_{K \times K} \end{pmatrix} \nonumber\\
    &= \lc\begin{pmatrix} \inner{\bs S,\bs S}_F  &\inner{\bs S,\bs P}_F\\\hline \inner{\bs P,\bs S}_F& \inner{\bs P,\bs P}_F \end{pmatrix} = \inner{\bs \sigma, \bs \sigma}_F.
\end{align}
\end{proof}
\end{prop}

The proof above hinges on the fact that $\bs T$ exists, which we later prove in Lemma \ref{prop:Texists}.

Similarly to the KW duality, we can also show that identity generators are in one-to-one correspondence to the dual symmetries. That is,
\begin{prop}$\ker \bs \epsilon_F=\bs \sigma_F \ker \tbs \sigma $ \label{prop:JWconstraint1}\end{prop}
\begin{prop}$\ker \tbs \epsilon=\tbs \sigma \ker \bs \sigma_F$
\label{prop:JWconstraint2} \end{prop}
Again, we defer the proofs to Appendix \ref{app:KWJWconstraints}. A summary of the relevant operators under the duality is listed in Table \ref{tab:JWsummary}. For completeness, the generating and excitation maps are listed below
\begin{align}
   \bs  \sigma_F &= (\bs S \ \bs P), &\tbs  \sigma &= (\tbs X \ \tbs Z) = \lc\begin{pmatrix}  \bs T_{N \times N}& \inner{\boldsymbol S,\boldsymbol P}_F \\\hline \mathbbm 1_{N\times N} & 0_{N\times K}\end{pmatrix}, \nonumber\\
   \bs \epsilon_F &= \bs \sigma_F^\dagger , & \tbs \epsilon &=\tbs \sigma^\dagger  = \begin{pmatrix}[c!{\color{\linecolor}\vrule} c] \mathbbm 1_{N\times N} & \bs T^\dagger_{N\times N}\\
  0_{K\times N} &\inner{\boldsymbol P,\boldsymbol S}_F \end{pmatrix}.
\end{align}

\subsubsection{Example: 1D JW duality}
As an example, let us bosonize the 1D fermionic Hamiltonian \eqref{equ:Majoranachain}. Since $\inner{\bs S, \bs S}_F = 0$, we can trivially choose the transmutation matrix $\bs T$ to be zero. Using Eq. \eqref{equ:XZforJW}, the dual to the hopping and fermion parity operators are
\begin{align}
   \tbs X &=\lc\begin{pmatrix}0 \\\hline 1\end{pmatrix}, & \tbs Z &=\lc\begin{pmatrix} 1+ x\\\hline 0 \end{pmatrix}.
\end{align}
We notice that this is precisely the same generators of the 1D Ising model. To conclude, the dual spin Hamiltonian is given by
\begin{equation}
    \tilde H=\frac{1}{2}\sum_{i}(\Delta+t) X_i  + (\Delta-t) Z_{i-1}X_iZ_{i+1} +\mu  Z_i Z_{i+1}.
    \label{equ:Majoranachaindual}
\end{equation}

Let us also demonstrate the correspondence between the identity generators and symmetries. The identity generator on the fermion side in Eq. \eqref{equ:identityrelation1D} dualizes to
\begin{align}
    \ker \tbs \epsilon = \tbs \sigma \ker \bs \sigma_F = \lc\begin{pmatrix}0 & 1+x \\\hline 1&0\end{pmatrix} \begin{pmatrix} \sum_i x^i\\\sum_i x^i\end{pmatrix} = \lc\begin{pmatrix} 0\\ \hline \sum_i x^i\end{pmatrix},
\end{align}
which is the global $\mathbb Z_2$ symmetry of the Ising model. On the other hand, the identity generator of the 1D Ising model dualizes to
\begin{align}
    \ker \bs \epsilon_F = \bs \sigma_F \ker \tbs \sigma = \lc\begin{pmatrix}1 & 1 \\\hline \bar x&1\end{pmatrix} \begin{pmatrix} \sum_i x^i\\0\end{pmatrix} = \lc \begin{pmatrix} \sum_i x^i\\ \hline \sum_i x^i\end{pmatrix},
\end{align}
which is the global fermion parity.

We remark that in this JW duality, the trivial phase of fermions ($\mu > 0$, $\Delta=t=0$) is dual to the ferromagnetic phase, while the ``Majorana'' chain ($\Delta=t$, $\mu=0$) maps to the paramagnet fixed point. This mapping is the opposite of the duality usually discussed in 1D, since the Majorana edge mode is often associated to the ground state degeneracy due to spontaneous symmetry breaking. We present two comments regarding this issue.

First, the duality presented has symmetry constraints at the level of states, meaning that we only allowed to map parity even fermionic states to $\mathbb Z_2$ symmetric states and vice versa. Therefore, only the symmetric combinations map to each other and there is no degeneracy in this restricted Hilbert space.

Second, the two different JW dualities differ precisely by an additional KW duality. However, as we will see in higher dimensions, this extra step is not always possible if $\bs T\ne 0$ since $\tbs X$ is no longer a transverse field. (This can also be seen as an indication of an 't Hooft anomaly). Therefore, the JW duality presented here is the natural one to generalize.

\subsection{Choices of the Interacting Terms and Transmutation Matrix}\label{JWproofs}
We will now turn to discuss the nuances of the JW duality. The first main difference from the KW duality is the necessity of the transmutation matrix $\bs T$. First, we must show that for any given choice of interaction terms $\bs S$, we can always construct such a $\bs T$. The proof is actually constructive. Denote $\bs S_j$ as the $j^\text{th}$ column of $\bs S$. Since $\inner{\bs S, \bs S}_F$ is Hermitian, we can explicitly construct $\bs T$ as an upper triangular matrix, keeping only entries $\inner{\bs S_j, \bs S_k}_F$ for $j<k$. The diagonal elements $\bs T_{jj}$ can be constructed by picking ``half'' the entries in $\inner{\bs S_j, \bs S_j}_F$.

The argument above is formally shown below.

\begin{lemma}\label{prop:Texists}
Given $\bs S$, there always exists a matrix $\bs T$ such that $\bs T+ \bs T^\dagger = \inner{\bs S, \bs S}_F$
\begin{proof}
Since the diagonal elements $\inner{\bs S_j, \bs S_j}_F$ has a unique decomposition
\begin{equation}
    \inner{\bs S_j, \bs S_j}_F = \sum_{\bs i} c^{(j)}_{\bs i}x_{1}^{i_1} \cdots x_{d}^{i_d}
\end{equation}
for coefficients $c^{(j)}_{\bs i} \in \mathbb F_2$, and $\inner{\bs S_j, \bs S_j}_F^\dagger = \inner{\bs S_j, \bs S_j}_F$, it follows that $c^{(j)}_{\bs i} = c^{(j)}_{-\bs i}$, where $-\bs i =(-i_1, \cdots ,-i_d)$. In particular, $c^{(j)}_{(0, \cdots ,0)}=0$. Therefore, define
\begin{align}
b^{(j)}_{\bs i} = \begin{cases}
c^{(j)}_{\bs i},& \parbox[t]{.4\columnwidth}{if the smallest $l$ where $i_l\ne 0$ satisfies $i_l>0$}\\
0,& \text{else}
\end{cases}. 
\end{align}
Then  $\bs T$ defined as
\begin{equation}
\label{equ:T}
\bs T_{jk} = \begin{cases}
\inner{\bs S_j, \bs S_k}_F,& j<k\\
\sum_{\bs i} b^{(j)}_{\bs i}x_{1}^{i_1} \cdots x_{d}^{i_d},& j=k\\
0,& j>k
\end{cases}
\end{equation}
has the desired property.
\end{proof}
\end{lemma}

The choice of $\bs T$ constructed above is not unique, however, we show in the following that such an ambiguity is not important.
\begin{prop}\label{choiceofT} 1. All choices of $\bs T$ differ by a Hermitian matrix\footnote{More correctly, anti-hermitian when generalizing to parafermions with $\mathbb F_p$} $\bs A = \bs A^\dagger$. \\
2. Different choices of $\bs T$ give rise to dual symmetries that are related by a basis transformation.

\begin{proof}
1. Let $\bs T$ and $\bs T'$ be two valid choices, and $\bs A =\bs T' + \bs T$. Then $\bs A+\bs A^\dagger = (\bs T+\bs T^\dagger) + (\bs T' + \bs T'^\dagger) =0$.\\
2. Since $\tbs \epsilon$ has the form
\begin{align}
  \tbs \epsilon & =  \begin{pmatrix}[c!{\color{\linecolor}\vrule} c] \mathbbm 1_{N\times N} & \bs T^\dagger_{N\times N}\\
  0_{K\times N} &\inner{\boldsymbol P,\boldsymbol S}_F \end{pmatrix},
\end{align}
one can verify that
\begin{equation}
    \tbs \epsilon' = \tbs \epsilon \bs U,
\end{equation}
where $\bs U=\begin{pmatrix} \mathbbm 1&\bs A \\ 0 &\mathbbm 1 \end{pmatrix}$ is a symplectic transformation. Therefore, the two symmetries are related by a basis transformation. Explicitly,
\begin{equation}
\ker \tbs \epsilon = \bs U \ker \tbs \epsilon'.
\end{equation}
\end{proof}
\end{prop}

Lastly,  we discuss the effects of modifying the interaction terms $\bs S$. For example, in the 1D Hamiltonian \eqref{equ:Majoranachain}, we chose the hopping operator $i\gamma_{j-1}'\gamma_{j}$ to be one of the generators. However, one could have instead chosen $i\gamma_{j-1}\gamma_{j}'$ as a generating operator, which will change $\bs S$ and its dual $\tbs X$.  Although that is indeed the case, we remark that all such choices must differ by a product of local fermion parity operators, and we can therefore show that this choice does not affect the dual symmetry in the Pauli Hamiltonian. Below, we implicitly sum over the index $k$, which runs over the labels of sites in the unit cell.

\begin{prop}
 The dual symmetry $\ker \tbs \epsilon$ is invariant (up to a basis transformation) under $\bs S_i \rightarrow \bs S_i' =\bs S_i+ \bs P_k f_{ki}$, for any polynomial $f_{ki} \in R$.
\begin{proof}
Under this change, the commutation matrix of $\bs S$ is modified to
\begin{equation}
\inner{\bs S_i',\bs S_j'}_F=\inner{\bs S_i,\bs S_j}_F + \inner{\bs S_i,\bs P_k}_Ff_{kj} + \inner{\bs P_k,\bs S_j}_F\bar f_{ki} 
    \end{equation}
Correspondingly, the transmutation matrix $\bs T$ can be chosen as
\begin{equation}
\bs T_{ij} \rightarrow \bs T_{ij}' = \bs T_{ij}+\inner{\bs S_i,\bs P_k}_Ff_{kj}
\end{equation}
up to a Hermitian matrix, which by Prop. \ref{choiceofT}, will only change $\ker \tbs \epsilon$ by basis transformation. 
Explicitly, since
\begin{align}
  \tbs \epsilon & =  \begin{pmatrix}[c!{\color{\linecolor}\vrule} c] \mathbbm 1_{N\times N} & \bs T^\dagger_{N\times N}\\
  0_{K\times N} &\inner{\boldsymbol P,\boldsymbol S}_F \end{pmatrix}
\end{align}
and $\bs T \rightarrow \bs T'$ implements a row operation on $\tbs \epsilon$, its kernel remains invariant.
\end{proof}
\label{prop:dualsymmetryinvariant}
\end{prop}

\section{Examples}\label{Examples}
In this section, we will present examples for the JW dualities, starting by reviewing the 2D and 3D examples with global symmetry discussed in Refs. \onlinecite{ChenKapustinRadicevic2018,ChenKapustin2019}. We will then move on to discuss more exotic examples, such as those with higher-form or subsystem fermion parity. A summary of the dualities considered in this section can be found in Table \ref{tab:resultssummary}.

Following the discussion of Prop. \ref{prop:dualsymmetryinvariant}, we remind that the choice of the interaction terms is not canonical, but they do not effect the final dual symmetry. Nevertheless, we will explicitly write interaction terms that mutually commute when possible, implying that the dual model has excitations that are purely bosonic. On the other hand, we will also mention examples where $\inner{\bs S,\bs S}_F \ne 0$, meaning that we are unable to find a set of interaction terms that commute given the corresponding symmetry. In those cases, we are only able to prove for some dualities that no such commuting choice exists. These are the dualities considered in Secs. \ref{ex:Global2D}, \ref{ex:Global3D}, \ref{ex:Fibonacci2D}, and \ref{ex:Fibonacci3D}, and we provide proofs in Appendix \ref{app:anomalyproof}. 

\subsection{JW for Global Fermion parity in 2D}\label{ex:Global2D}
To warm up, let us write the JW duality in 2D reviewed in Sec. \ref{JWreview} in the algebraic formalism. Again, this should be compared to the 2D KW duality presented in Sec. \ref{KW2D}.

Given the Hamiltonian \eqref{equ:2Dglobalfermion}, the generating operators can be chosen as
\begin{align}
  \bs S & = \lc \begin{pmatrix}
    1 & 1 \\ \hline  x & y
    \end{pmatrix},&
  \bs P &=\lc  \begin{pmatrix}
    1 \\ \hline 1
    \end{pmatrix}.
\end{align}
Therefore, the generating and excitation maps are
\begin{align}
   \bs \sigma_F &= \lc\begin{pmatrix} 1 & 1 & 1 \\ \hline  x & y & 1\end{pmatrix}, &\bs \epsilon_F &=\begin{pmatrix}[c!{\color{\linecolor}\vrule} c] 1& \bar x \\
   1& \bar y\\
   1& 1\end{pmatrix}.
\end{align}
The global fermion parity symmetry $\prod_{i,j} P_{i,j}$ is given by
\begin{align}
\ker \bs \epsilon_F = \lc \begin{pmatrix} \sum_{ij}x^iy^j\\ \hline \sum_{ij}x^iy^j\end{pmatrix},
\end{align}
and the identity generators are
\begin{align}
   \ker \bs \sigma_F =  \begin{pmatrix}
   1+y &\sum_i x^i & 0\\
   1+x & 0 & \sum_j y^j\\
   x+y &\sum_i x^i & \sum_j y^j
   \end{pmatrix}.
\end{align}

To perform the JW duality, we first compute the commutation matrix
\begin{align}
\inner{\bs S,\bs S}_F=\begin{pmatrix} 0 & 1+\bar x y\\1+x\bar y &0 \end{pmatrix}
\end{align}
and choose a corresponding transmutation matrix as
\begin{align}
\bs T=\begin{pmatrix} 0 & \bar x y\\1 &0 \end{pmatrix}.
\end{align}
Here, it is clear that $\bs T + \bs T^\dagger = \inner{\bs S,\bs S}_F$. Therefore, the dual operators are
\begin{align}
   \tbs X &=\lc\begin{pmatrix}    
   0 & \bar x y \\
    1 & 0 \\
    \hline  
    1 &0\\
   0&1 \end{pmatrix}, &\tbs Z &=\lc\begin{pmatrix}  1+ \bar x\\
    1+ \bar y\\
    \hline
    0\\
    0\
    \end{pmatrix},
\end{align}
which correspond to the terms in the Hamiltonian \eqref{equ:JWdual2D}.
From Prop. \ref{prop:JWconstraint2}, the dual symmetries can be calculated to be
\begin{align}
   \ker \tbs \epsilon = \tbs \sigma \ker \bs \sigma_F =  \lc\begin{pmatrix}
   1+x &0 & \sum_i y^i\\
   x(1+\bar y) & \bar y \sum_i x^i & 0\\
   \hline
   1+y &\sum_i x^i & 0\\
   1+x & 0&\sum_i y^i
   \end{pmatrix},
\end{align}
which are precisely the anomalous 1-form symmetries in Eq. \eqref{equ:1formF2D}.

\begin{figure}
    \centering
    \includegraphics{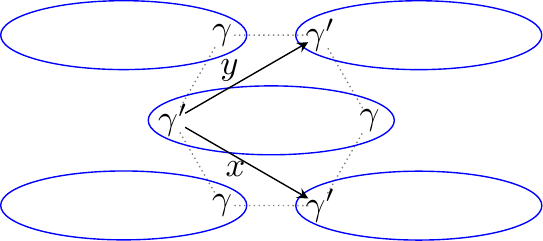}
    \caption{The stabilizer of the Majorana color code, written with the unit cell shown in blue. The translation vectors are denoted by $x$ and $y$.} 
    \label{fig:MCC}
\end{figure}

As an application, we can also dualize the Majorana color code\cite{VijayHsiehFu2015,VijayHaahFu2016}. With the choice of unit cell as given in Fig. \ref{fig:MCC}, the Hamiltonian is given by
\begin{align}
    H_{MCC} = -i\sum_i \includegraphicsr{1}{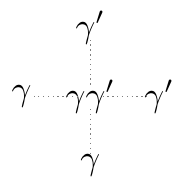}.
\end{align}
Algebraically, wee see that
\begin{align}
\bs H_{MCC}&= \lc\begin{pmatrix}
1+\bar x+\bar y\\
\hline1+ x +  y
\end{pmatrix} = \bs \sigma_F \begin{pmatrix}
1+\bar x\\
1+\bar y\\
1
\end{pmatrix},
\label{equ:MCC}
\end{align}
which means that it is a valid term in our generating Hamiltonian and can therefore be dualized. We remark that the stabilizer above alone has a larger symmetry, which preserves fermion parity in individual diagonal lines. Nevertheless, these symmetries  are explicitly broken by the interaction terms $\bs S$. The duality in the case of subsystem symmetry can be found in Sec. \ref{ex:2DSSPT2}.

Dualizing this stabilizer gives
\begin{align}
\tbs H_{MCC}&= \tbs \sigma \begin{pmatrix}
1+\bar x\\
1+\bar y\\
1
\end{pmatrix}=
\lc\begin{pmatrix}
1+x\bar y\\
\bar x + \bar y\\
\hline
1+ \bar x\\
1+\bar y
\end{pmatrix} = \includegraphicsr{1}{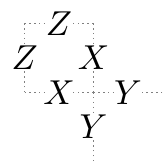}.
\end{align}
The ground state of this model under the gauge constraints exhibits a $\mathbb Z_2^2$ topological order. The corresponding Wilson loops are the nonlocal line operators in Eq. \eqref{equ:1formF2D} and the dual of the line operators of the Majorana color code. Their 't Hooft loops are obtained in a similar fashion.

It is interesting in its own right to analyze the stabilizer code
\begin{equation}
    H = - \sum_i  \includegraphicsr{1}{2DJW_1form1.pdf} + \includegraphicsr{1}{MCCdualglobal.pdf}.
\end{equation}
 This model can be thought of as the result of ``gauging'' the global fermion parity symmetry of the Majorana color code. Since this stabilizer has the same topological order as the (Pauli) color code\cite{BombinMartin-Delgado2006}, it would be interesting to compare their performances.

\subsection{JW for Global Fermion Parity in 3D} \label{ex:Global3D}
The same exercise can be done for a 3D cubic lattice\cite{ChenKapustin2019}. There are now three interaction terms
\begin{equation}
    \bs S = \lc\begin{pmatrix}
    1 & 1 &1 \\ \hline  x & y &z
    \end{pmatrix},
    \label{equ:S3D}
\end{equation}
and the corresponding commutation matrix is \begin{align}
\inner{\bs S,\bs S}_F=\begin{pmatrix} 0 & 1+\bar x y & 1+ \bar x z\\1+x\bar y &0 & 1+\bar y z\\
1+x\bar z& 1+ y \bar z&0
\end{pmatrix}.
\end{align}
We choose \footnote{in Fig. 2 of Ref. \onlinecite{ChenKapustin2019}, $\bs T = \begin{pmatrix} 0 & 1 & 1\\x\bar y &0 & 1\\ x\bar z&y\bar z&0 \end{pmatrix}$, which differs from our choice by a Hermitian matrix}
\begin{align}
\bs T=\begin{pmatrix} 0 & \bar x y & 1\\1 &0 & \bar y z\\ \bar z x &1&0 \end{pmatrix},
\end{align}
so that the result is invariant under a $C_3$ rotation around the $(1,1,1)$ axis. In the algebraic language, this means that it is invariant under the cyclic permutation of symbols $x\rightarrow y \rightarrow z \rightarrow x$ along with the rows and columns of $\bs T$.

The dual operators are
\begin{align}
   \tbs X &=\lc\begin{pmatrix} 0 & \bar x y & 1\\1 &0 & \bar y z\\ \bar z x &1&0\\
    \hline  
    1 &0&0\\ 0&1&0\\0&0&1 \end{pmatrix},
    & \tbs Z &=\lc\begin{pmatrix}  1+ \bar x\\
    1+ \bar y\\
    1+\bar z\\
    \hline
    0\\
    0\\
    0
    \end{pmatrix}.
\end{align}
There are three local and three nonlocal symmetry constraints. A similar calculation as the 2D case gives 
\begin{widetext}
\begin{align}
   \ker \tbs \epsilon = \tbs \sigma \ker \bs \sigma_F =  \lc\begin{pmatrix}
   1+z+\bar x z (1+y)&   z(1+\bar x) &1+x & 0 & \sum_i y^i & \bar x \sum_i z^i\\
     1+y   &   1+x+\bar y x (1+z) & x(1+\bar y)  & \bar y \sum_i x^i & 0 & \sum_i z^i \\
   y(1+\bar z) &1+z  & 1+y+\bar z y (1+x) & \sum_i x^i & \bar z \sum_i y^i &0\\
   \hline
0 &1+z & 1+y & \sum_i x^i & 0 & 0\\
1+z& 0 & 1+x& 0 & \sum_i y^i &0\\
1+y& 1+x &0 &0 & 0 & \sum_i z^i
   \end{pmatrix}.
   \label{equ:2formF}
\end{align}
\end{widetext}
The symmetries listed here are 2-form symmetries. When $\bs Z$ is the stabilizer, it is deconfined and the ground state has the same topological order as a ``twisted'' 3D toric code where the emergent point particle has fermionic statistics\cite{WalkerWang2012,KapustinThorngren2017}.

As an application, we dualize the six Majorana Hamiltonians with extensive ground state degeneracy proposed in Ref. \onlinecite{VijayHaahFu2016}, the first of which is the Majorana checkerboard model. Like the 2D case, these models alone actually have a larger symmetry, but we explicitly break them by adding the interaction terms $\bs S$.  The results are summarized in Table \ref{tab:JW3Dsummary} of Appendix \ref{app:Majoranaduals}. 

\subsection{1-form fermion parity in 3D}\label{ex:1form}
Lattice models with higher-form symmetry can be constructed for spin models\cite{KapustinThorngren2017,Yoshida2016,TsuiWen2020}\footnote{In some references, they are referred to as faithful/nonrelativistic n-symmetry as opposed to unfaithful/relativistic n-form symmetry considered in high energy}. For example, the KW dual of the 2D and 3D transverse-field Ising models have 1-form and 2-form symmetries respectively.

Here, we consider a 3D fermionic model with 1-form fermion parity symmetry. We place fermions on the edges of a cubic lattice and consider interaction terms to be generated by Majorana operators of four links surrounding a plaquette as shown in Table \ref{tab:1form}. The local 1-form symmetries are generated by a product of six fermion parity operators on links surrounding a vertex, and the nonlocal symmetries form nontrivial 2-cycles around the torus in the dual lattice.

In the algebraic notation, we associate each link to a vertex, so that we have a cubic lattice with three sites per unit cell. We can write
\begin{align}
    \bs S &= \lc\begin{pmatrix}
    0 & 1 & 1 \\
    1 & 0 & 1\\
    1 & 1 & 0\\
    \hline  0&z&y\\
    z&0&x\\
    y&x&0
    \end{pmatrix}, &
        \bs P &= \lc\begin{pmatrix}
    1&0&0 \\
    0&1&0\\
    0&0&1\\
    \hline     
    1&0&0 \\
    0&1&0\\
    0&0&1\\
    \end{pmatrix}.
\end{align}
The 1-form symmetries are generated by
\begin{equation}
    \ker \bs \epsilon_F = 
\lc\begin{pmatrix}
    1+\bar x&\sum_{ij}y^iz^j&0 &0 \\
    1+\bar y&0 & \sum_{ij}x^iz^j &0\\
    1+\bar z&0&0 & \sum_{ij}x^iy^j\\
    \hline     
    1+\bar x&\sum_{ij}y^iz^j&0 &0 \\
    1+\bar y&0 & \sum_{ij}x^iz^j &0\\
    1+\bar z&0&0 & \sum_{ij}x^iy^j
    \end{pmatrix}.   
\end{equation}
We choose the following transmutation matrix \begin{equation}
    \bs T = 
\begin{pmatrix}
  0 & x\bar y & 1 \\
  1 & 0 & y\bar z\\
z\bar x & 1 & 0
    \end{pmatrix}.     
\end{equation}
Therefore, the dual operators are
\begin{align}
    \tbs X &= \lc\begin{pmatrix}
  0 & x\bar y & 1 \\
  1 & 0 & y\bar z\\
z\bar x & 1 & 0\\
    \hline  1&0&0\\
    0&1&0\\
    0&0&1
    \end{pmatrix}, &
        \bs Z &= \begin{pmatrix}
    0&1+\bar z&1+\bar y \\
    1+\bar z&0&1+\bar x\\
   1+\bar y&1+\bar x&0\\
    \hline     
    0&0&0 \\
    0&0&0\\
    0&0&0
    \end{pmatrix}.
\end{align}
The duality is depicted visually in Table \ref{tab:1form}. In the dual model, there are local symmetry constraints coming from the following local identity generator
\begin{align}
    \tbs G &= \tbs \sigma \begin{pmatrix}
    1+x\\1+y\\1+z\\y+z\\x+z\\x+y
    \end{pmatrix} = \begin{pmatrix}
    1+x+y+x\bar z\\1+y+z+y\bar x\\  1+z+x+z\bar y\\ \hline1+x\\1+y\\1+z
    \end{pmatrix}\\
    &=\includegraphicsr{1}{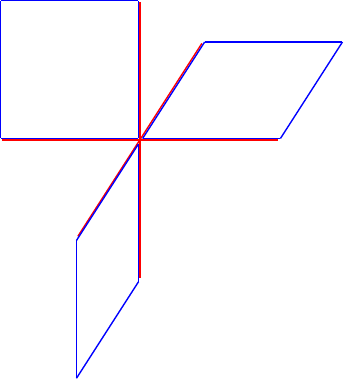}
    \label{equ:1formanomalous?}
\end{align}

The ground state of this dual spin model when $\bs Z$ dominates is actually just the 3D toric code. This is because the local symmetry constraint can be written as the vertex terms attached with one plaquette term per orientation as shown above.

It is not clear whether this 1-form symmetry is anomalous. If it is so, we conjecture from the bulk-boundary correspondence that the anomaly should be matched by a bulk SPT with $\mathbb Z_2$ 1-form symmetry. Such SPT in 4+1D is classified by $\mathbb Z_2$\cite{ZhuLanWen2019,WanWang2018}, with response to a background $\mathbb Z_2$ 2-form $B$ given by $B \cup Sq^1B$, where $Sq^1$ is the first Steenrod square\cite{Steenrod1947}.

\begin{table}[t]
\caption{Duality of operators (up to a sign) for a fermion system with 1-form symmetry in 3D to a spin system with $\mathbb Z_2$ 1-form symmetry. The interaction terms living on plaquettes are sent to the red edges in the dual lattice. Here, the blue and red lines represent Pauli $Z$'s and $X$'s, respectively.}
\begin{tabular}{|c|c|c|c|}
\hline
Fermion  & Spin & Fermion & Spin \\
\hline
\includegraphicsr{1}{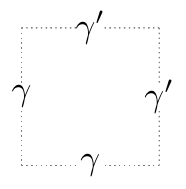}  & \includegraphicsr{1}{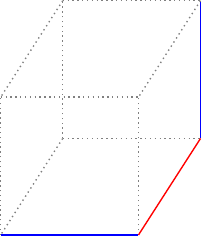} & \includegraphicsr{1}{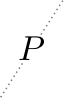}  & \includegraphicsr{1}{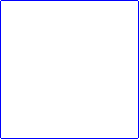} \\
\hline
\includegraphicsr{1}{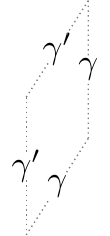}  & \includegraphicsr{1}{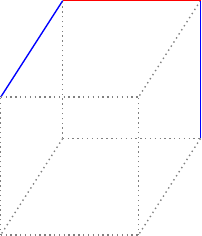} & \includegraphicsr{1}{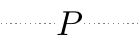}  & \includegraphicsr{1}{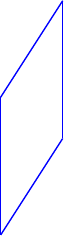} \\
\hline
\includegraphicsr{1}{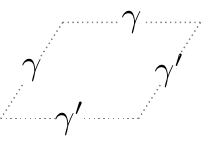}  & \includegraphicsr{1}{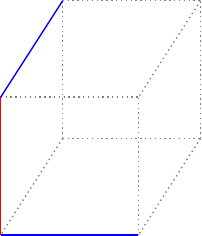} &\includegraphicsr{1}{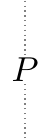}  & \includegraphicsr{1}{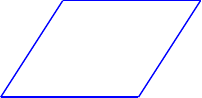}\\
\hline
\end{tabular}
\label{tab:1form}
\end{table}

\begin{table*}[t]
\caption{Duality of operators (up to a sign) for a fermion system with  line symmetry to a spin system with $\mathbb Z_2$  line symmetry.}
\begin{tabular}{|c|c|c|c|}
\hline
Fermion & Comment & Spin & Comment\\
\hline
$i \gamma \gamma'$  & Trivial & $\begin{array}{cc} 
 Z & Z\\
Z &Z
\end{array}$ & Symmetry breaking\\
\hline
$\begin{array}{cc}
  \gamma' & \gamma\\
 \gamma' &\gamma
\end{array}$  & SSB in $y$ direction & $\begin{array}{ccc}
 & & \\
&X &\\
 & & \\
\end{array}$ & Trivial\\
\hline
$\begin{array}{cc}
 \gamma & \gamma\\
 \gamma' & \gamma'
\end{array}$ & SSB in $x$ direction & $\begin{array}{ccc}
Z &Z & \\
Z&X &Z\\
 & Z& Z\\
\end{array}$ & SSPT (2D cluster state)\\
\hline
$\begin{array}{ccc}
  \gamma' & \gamma &\\
 \gamma' & &\gamma\\
 & \gamma' &\gamma
\end{array}$ & $\mathbb Z_2$ Topological order 
&
$\begin{array}{cc}
Y & Z\\
Z & Y
\end{array}$ & $\mathbb Z_2$ Topological order (Wen plaquette)\\
\hline
\end{tabular}
\label{tab:2DSSPT}
\end{table*}
\begin{table*}[t]
\caption{Duality of operators (up to a sign) for a fermion system with (diagonal) line symmetry to a  spin system with $\mathbb Z_2$ (diagonal) line symmetry.}
\begin{tabular}{|c|c|c|c|}
\hline
Fermion & Comment & Spin & Comment\\
\hline
$i \gamma \gamma'$  & Trivial & $\begin{array}{ccc} 
& Z & \\
Z &&Z\\
&Z&
\end{array}$ & Symmetry breaking\\
\hline
$\begin{array}{ccc}
 & \gamma &\\
  \gamma' & & \gamma \\
 &\gamma' &\\
\end{array}$  & Symmetry breaking  & $\begin{array}{ccc}
 & & \\
&X &\\
 & & \\
\end{array}$ & Trivial\\
\hline
$\begin{array}{ccc}
 & \gamma &\\
  \gamma & & \gamma' \\
 &\gamma' &\\
\end{array}$  & Symmetry breaking  & $\begin{array}{ccccc}
 &Z& &Z& \\
Z&&X &&Z\\
 &Z& &Z& \\
\end{array}$ & $\mathbb Z_2$ SSPT\\
\hline
$\begin{array}{ccc}
 & \gamma &\\
 \gamma' &  i\gamma \gamma' & \gamma \\
 &\gamma' &\\
\end{array}$ & $\mathbb Z_2$ Topological order (Majorana color code)
&
$\begin{array}{ccc}
 & Z &\\
 Z & X& Z \\
 &Z &\\
\end{array}$ & $\mathbb Z_2$ SSPT \\
\hline
\end{tabular}
\label{tab:2DSSPT2}
\end{table*}

\subsection{Subsystem Fermion parity in 2D}\label{ex:line2D_1}
The prototypical example of a 2D system with line subsystem symmetry is the plaquette Ising a.k.a. the Xu-Moore model\cite{XuMoore2004}. Such models with subsystem symmetry can host subsymmetry-broken phases, subsystem symmetry-protected phases\cite{Youetal2018,DevakulWilliamsonYou2018}, or topological order\cite{Wen2003,TantivasadakarnVijay2019}. Here, we consider an analogous fermionic system, which will turn out to be JW dual to such a spin system with line symmetry.

Consider a square lattice with interaction term
\begin{equation}S=\begin{array}{cc}
  \gamma' & \gamma \\
 \gamma' & \gamma\\
\end{array},\end{equation} or in the algebraic notation,
\begin{align}
\bs S&= \lc\begin{pmatrix}
x(1+y) \\
\hline 1+y
\end{pmatrix}.
\end{align}
The Hamiltonian has subsystem fermion parity symmetry, defined as the product of the local fermion parity operators on each individual vertical and horizontal lines. They correspond to
\begin{align}
\ker \tbs \epsilon_F&= \lc\begin{pmatrix}
\sum_i x^i & \sum_i y^i\\
\hline 
\sum_i x^i & \sum_i y^i
\end{pmatrix}.
\end{align}
To perform the JW duality, we compute
\begin{align}
\inner{\bs S,\bs S}_F =0,
\end{align}
that is, all the interaction terms commute, and so we can trivially choose $\bs T =0$. The dual operators are therefore
\begin{align}
\tbs X&= \lc\begin{pmatrix}
0\\
\hline 
1
\end{pmatrix}, &
\tbs Z&= \lc\begin{pmatrix}
(1+\bar x) (1+\bar y)\\
\hline 
0
\end{pmatrix},
\end{align}
and the dual symmetry is
\begin{align}
\ker \tbs \epsilon= \ker \begin{pmatrix}[c!{\color{\linecolor}\vrule} c]
1 & 0\\
0 & (1+x)(1+y)
\end{pmatrix} =
\lc\begin{pmatrix}
0 & 0\\
\hline 
\sum_i x^i & \sum_i y^i 
\end{pmatrix}.
\end{align}
Thus, we see that fermion parity and the interaction term, respectively, map to the Ising term and the transverse field in the Xu-Moore model, protected by horizontal and vertical line subsystem symmetries. Pictorially,
\begin{align}
    i \gamma \gamma'  &\rightarrow \begin{array}{cc} 
 Z & Z\\
Z &Z
\end{array}, &
\begin{array}{cc}
  \gamma' & \gamma\\
 \gamma' &\gamma
\end{array} &\rightarrow \begin{array}{ccc}
 & & \\
&X &\\
 & & \\
\end{array}.
\label{equ:2Dlinemapping}
\end{align}

Since the interaction term fully commutes, the Hamiltonian consisting only of this term is exactly solvable. However, it is symmetry breaking. On a torus of size $L_x \times L_y$, the product of $S$ along any column vanishes. More precisely,
\begin{equation}
    \sum_i y^i \bs S =0.
\end{equation}
Hence, the stabilizer has an extensive ground state degeneracy of $L_x$, which can be labeled by the eigenvalues of the $L_x$ vertical line symmetries. The degeneracy can be broken by explicitly breaking the vertical symmetries with $\begin{array}{cc} \gamma' & \gamma \end{array}$, which commutes with $S$. The model is still exactly solvable, with remaining horizontal symmetries, and the ground states are decoupled horizontal Majorana chains.

Now, consider the following operator, which is a $90^\circ$ rotated interaction term
\begin{equation}
\begin{array}{cc}
\gamma & \gamma\\
\gamma' &\gamma'
\end{array}.
\end{equation}
A Hamiltonian consisting of only this term spontaneously breaks the horizontal line symmetries. Interestingly, the result of bosonizing this operator using the map \eqref{equ:2Dlinemapping} is a 2D cluster state, which is the stabilizer for the $\mathbb Z_2$ SSPT phase\cite{Youetal2018,DevakulWilliamsonYou2018} given by
\begin{equation}
\begin{array}{ccc}
Z &Z & \\
Z&X &Z\\
 & Z& Z\\
\end{array},
\end{equation}
Furthermore, one can also consider the KW dual of the above stabilizer, which is the Wen-plaquette model\cite{Wen2003}
\begin{equation}
\begin{array}{cc}
Y & Z\\
Z & Y
\end{array}.
\end{equation}
The ground state of this Hamiltonian spontaneously breaks the $\mathbb Z_2$ line symmetry and is distinct from the symmetry broken phase in the Xu-Moore model\cite{TantivasadakarnVijay2019}. The JW dual of this stabilizer is
\begin{equation}
\begin{array}{ccc}
  \gamma' & \gamma &\\
 \gamma' & &\gamma\\
 & \gamma' &\gamma
\end{array}.
\end{equation}
Such a model is reminiscent of the Majorana color code\cite{VijayHsiehFu2015}. We can calculate the ground state degeneracy of this model by a similar counting argument. First, the stabilizer above can be written algebraically as the vector
\begin{align}
\lc
    \begin{pmatrix}
    x+y+x\bar y\\
    \hline
    \bar x+\bar y+ \bar x y
    \end{pmatrix}.
\end{align}
Next, we notice that
\begin{align}
\lc
  \sum_{ij} x^iy^{3j}  \begin{pmatrix}
    x+y+x\bar y\\
    \hline
    \bar x+\bar y+ \bar x y
    \end{pmatrix} = \sum_{ij}x^iy^j\begin{pmatrix}
    1\\
    \hline
    1
    \end{pmatrix} = \bs P \sum_{ij}x^iy^j.
\end{align}
That is, the product of these stabilizers on every three rows is the global fermion parity. Therefore, the operators can be tripartited so that the product of all operators in each partition is the global fermion parity. By counting the number of eigenvalues each operator fixes, we find that on a torus with $L_xL_y$ complex fermions ($2L_xL_y$ Majorana fermions), the ground state degeneracy is
\begin{align}
    2^{L_xL_y-1}/(2^{L_xL_y/3-1})^3 = 4.
\end{align}
Therefore, the JW dual of the Wen-plaquette model also realizes a $\mathbb Z_2$ topological order.

A summary of the fermionic operators and their duals is given in Table \ref{tab:2DSSPT}. The fact that the interaction terms all commute allows us to further perform a KW duality on the spin system. In Appendix \ref{app:naiveJW}, we show that the combined duality is actually a ``naive'' JW duality that one would do in a 2D system.

\subsection{$\mathbb Z_2^F$ line symmetry in diagonal directions in 2D}\label{ex:2DSSPT2}
Consider a square lattice with interaction term 
\begin{equation}\begin{array}{ccc}
 & \gamma &\\
  \gamma' & & \gamma \\
 &\gamma' &\\
\end{array},\end{equation} or
\begin{align}
\bs S&= \lc\begin{pmatrix}
x+y \\
\hline \bar x +\bar y
\end{pmatrix}
\end{align}
in the algebraic notation.

The symmetries in this system are diagonal line symmetries, with normals pointing in the $(1,1)$ and $(1,-1)$ directions
\begin{align}
\ker \epsilon_F&= \lc\begin{pmatrix}
\sum_i (xy)^i & \sum_i (x\bar y)^i\\
\hline 
\sum_i (xy)^i & \sum_i (x\bar y)^i
\end{pmatrix}.
\end{align}
Alternatively, by enlarging the unit cell to two sites, it can be viewed as $\mathbb Z_2^F \times \mathbb Z_2^F$ line symmetries in the vertical and horizontal directions as in the previous subsection. Again, because all interaction terms commute, we can dualize to spin operators

\begin{align}
\tbs X&= \lc\begin{pmatrix}
0\\
\hline 
1
\end{pmatrix}, &
\tbs Z&= \lc\begin{pmatrix}
x+y + \bar x + \bar y\\
\hline 
0
\end{pmatrix},
\end{align}
with dual line symmetries also in the same directions.

In this duality, we can consider dualizing the Majorana color code
\begin{align}
\bs H_{MCC}&= \lc \begin{pmatrix}
1+x+y\\
\hline1+\bar x + \bar y
\end{pmatrix} =\bs \sigma_F\begin{pmatrix}
1\\
1
\end{pmatrix}.
\end{align}
Note that here we have inverted the model compared to Eq. \eqref{equ:MCC}. The dual stabilizer is given by
\begin{align}
\tbs H_{MCC}&= \tbs \sigma \begin{pmatrix}
1\\
1
\end{pmatrix} = \lc \begin{pmatrix}
x+y + \bar x + \bar y\\
\hline1
\end{pmatrix},
\end{align}
which is just the 2D cluster state, an SSPT protected by the dual line symmetries.

One can also consider bosonizing the $90^\circ$ rotated interaction term. This turns out to be a different SSPT. The mapping of operators in this duality is summarized in Table \ref{tab:2DSSPT2}.

\subsection{(100) Fermion Planar symmetry in a cubic lattice}\label{ex:planar3D_1}
The plaquette Ising model for a cubic lattice in 3D has a KW dual which is a gauge theory. In the deconfined phase, the ground state is the same as that of the X-cube model\cite{VijayHaahFu2016,Radicevic2019}. The gauge constraints are the ``cross'' terms in the X-cube model which, when enforced energetically, forbids lineon excitations.
We will now consider the fermionic analog of this model and perform a JW duality. Consider the Hamiltonian
\begin{align}
 H = -\sum_{i,j,k} &\left [ \gamma_{i,j,k}\gamma_{i,j+1,k} \right. \gamma_{i,j,k+1} \gamma_{i,j+1,k+1} \nonumber\\
 &+\gamma_{i,j,k}\gamma_{i+1,j,k} \gamma_{i+1,j,k+1} \gamma_{i,j,k+1}\nonumber\\
& +\gamma_{i,j,k}\gamma_{i+1,j,k} \gamma_{i,j+1,k} \gamma_{i+1,j+1,k} + \mu P_{i,j,k}\left. \right]\nonumber\\
 = -\sum_{i,j,k} &\left [ \includegraphicsr{0.5}{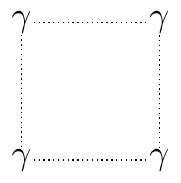}+ \includegraphicsr{0.5}{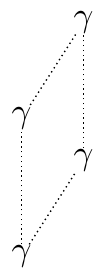}+ \includegraphicsr{0.5}{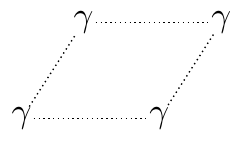} + \mu P_{i,j,k} \right],
\end{align}
which has four-body interaction terms consisting of four Majorana operators at the corners of each face of the cubic lattice. Algebraically,
\begin{equation}
    \bs S = \lc \begin{pmatrix} (1+y)(1+z) & (1+x)(1+z) & (1+x)(1+y) \\\hline 0&0&0 \end{pmatrix}.
\end{equation}
The symmetries of this model are fermion parity conservation in each individual $xy$, $yz$, and $xz$ planes
\begin{align}
    \ker \bs \epsilon_F =  \lc \begin{pmatrix}
      \sum_{ij} x^iy^j&\sum_{ij} y^iz^j&\sum_{ij} z^ix^j\\ \hline \sum_{ij} x^iy^j&\sum_{ij} y^iz^j&\sum_{ij} z^ix^j
    \end{pmatrix},
\end{align}
and the identity generators are
\begin{widetext}
\begin{align}
    \ker \bs \sigma_F =\begin{pmatrix}
      1+x & 0 & \sum_{i}y^i&\sum_{i}z^i&0&0&0&0\\
      1+y & 1+y &0&0&\sum_{i}x^i&\sum_{i}z^i&0&0\\
      0&1+z &0&0&0&0&\sum_{i}x^i&\sum_{i}y^i\\
      0&0&0&0&0&0&0&0
    \end{pmatrix}.
\end{align}
The first two are locally generated from a product of four plaquettes forming a ``belt'' around a cube. The last six are various nonlocal products of a ``belt'' of plaquettes wrapping around the torus.

To perform the duality, we first calculate the commutation matrix
\begin{equation}
    \inner{\bs S,\bs S}_F = \begin{pmatrix}
(y +\bar y) (z+\bar z) & (1+ x)(1+\bar y) (z+\bar z) & (1+ x)(y +\bar y) (1+\bar z)\\
(1+\bar x)(1+ y) (z+\bar z) & (x +\bar x) (z+\bar z) &  (x+\bar x)(1 + y)(1 +\bar z)\\
(1+ \bar x)(y +\bar  y) (1+ z) &(x+\bar x)(1 +\bar  y)(1 + z)& (x +\bar x) (y+\bar y)
\end{pmatrix} .
\end{equation}
We choose the following transmutation matrix
\begin{equation}
\bs T= \begin{pmatrix}
\bar y (z+\bar z) & (1+ x)(1+\bar y) \bar z & (1+ x)\bar y (1+ \bar z)\\
(1+\bar x)(1+  y) \bar z & (x +\bar x) \bar z &  \bar x(1 + y)(1 +\bar z)\\
(1+  \bar x)\bar  y (1+  z) &\bar x(1 +\bar  y)(1 + z)& \bar x (y+\bar y)
\end{pmatrix},
\end{equation}
so that it is invariant under a $C_3$ rotation around the $(1,1,1)$ axis on the cubic lattice. Namely, a cyclic permutation of the rows and columns and the coordinates $x \rightarrow y \rightarrow z \rightarrow x$ leaves $\bs T$ invariant.

To summarize, the JW dual has operators
\begin{align}
   \tbs X &=\lc \begin{pmatrix}\bar y (z+\bar z) & (1+ x)(1+\bar y) \bar z & (1+ x)\bar y (1+ \bar z)\\
(1+\bar x)(1+  y) \bar z & (x +\bar x) \bar z &  \bar x(1 + y)(1 +\bar z)\\
(1+  \bar x)\bar  y (1+  z) &\bar x(1 +\bar  y)(1 + z)& \bar x (y+\bar y) \\ \hline 1&0&0\\0&1&0\\0&0&1\end{pmatrix},
 &\tbs Z &=\lc\begin{pmatrix} (1+\bar y)(1+\bar z) \\ (1+\bar x)(1+\bar z) \\ (1+\bar x)(1+\bar y)\\
 \hline0\\0\\0\end{pmatrix}.
\end{align}
The symmetries of the dual model includes both local and nonlocal constraints. For simplicity, let us only write down the local gauge constraints $\tbs G \subset \ker \tbs \epsilon$ obtained by dualizing the local identity generators (the first two columns of $\ker \bs \sigma_F$).
\begin{equation}
\tbs G  = \lc\begin{pmatrix}
(1+x)(y\bar z + \bar y z) & (1+x)(y\bar z + \bar y z)\\
0 & (1+y)(x\bar z + \bar x z) \\
(1+z)(x\bar y + \bar x y) & 0 \\
\hline
1+x & 0\\
1+y & 1+y\\
0 & 1+z\\
\end{pmatrix}.
\end{equation}
Pictorially, the dual spin model on the dual lattice can be written as
\begin{equation}
   \tilde H = -\sum \left [  \includegraphicsr{0.5}{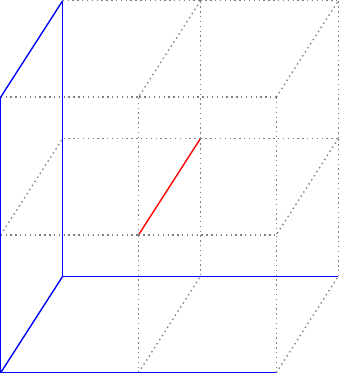} +\includegraphicsr{0.5}{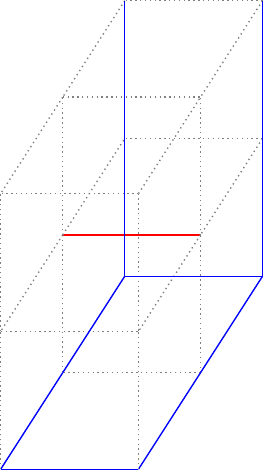} + \includegraphicsr{0.5}{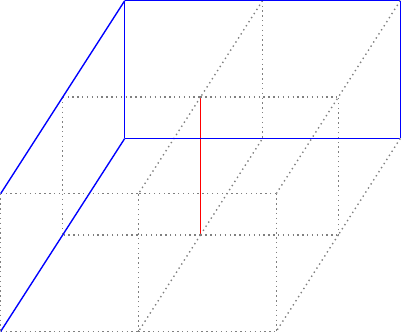} +\mu \includegraphicsr{0.5}{Xcube1.pdf} \right],
\end{equation}
with local gauge constraints
\begin{align}
\raisebox{-.5\height}{\includegraphics[scale=0.5]{Xcubeconstraint1.pdf}}=\raisebox{-.5\height}{\includegraphics[scale=0.5]{Xcubeconstraint2.pdf}}=\raisebox{-.5\height}{\includegraphics[scale=0.5]{Xcubeconstraint3.pdf}}=1.
\end{align}
\end{widetext}
Here, the red and blue lines denote Pauli $X$ and $Z$ operators, respectively, and we have drawn the third gauge constraint, which is the product of the first two constraints for rotational symmetry.

When $\mu \gg 1$, the model is deconfined and exactly solvable. The emergent excitations from violating the cube terms ($\tbs Z$) are fractons. Furthermore, because the gauge constraints are modified ``cross'' terms of the X-cube model, the mobility of these fractons are exactly identical: Four fractons can be created using the dual of the interaction terms ($\tbs X$), and pairs of fractons can move in a plane. The ``twisted'' X-cube model given in Fig. \ref{fig:twistedXcube} is a stabilizer code obtained from combining the cube term $\tbs Z$ and the local symmetry constraints $\tbs G$ of this dual Ising model.

Under this duality, we can also dualize the Majorana model\cite{VijayHaahFu2015} given by  $\bs H_{M3} = \lc \begin{pmatrix} f_3 \\ \hline \bar f_3\end{pmatrix}$, where $f_3= 1+x+y+yz+xz$, since it can be written as $\bs H_{M3} = \bs \sigma_F \bs g$, for the generator label
\begin{equation}
  \bs g =  \begin{pmatrix} 1+\bar y \bar z\\1+\bar x \bar z \\0\\1+(\bar x+\bar y)(1+\bar z) \end{pmatrix}.
\end{equation}
The resulting dual Hamiltonian is
\begin{equation}
  \tbs H_{M3} = \tbs \sigma \bs g = \lc\begin{pmatrix} 
  1 + \bar x + \bar x \bar y + \bar z^2+ x \bar z + \bar y \bar z + x \bar y \bar z + \bar y z\\
  1 + \bar y + \bar x \bar y + \bar x^2+ y \bar z + \bar x \bar z + y \bar x \bar z + \bar z x\\
  1 + \bar x +\bar y +\bar x \bar y +  \bar x \bar z + \bar y \bar z + \bar x z + \bar y z\\
  \hline
1+\bar y \bar z\\1+\bar x \bar z \\0\end{pmatrix}.
\end{equation}
As fermion models, it is known that the Hamiltonians generated by $\bs H_{M3}$ and $\bs P$ realize different fracton phases since $\bs H_{M3}$ has an extensive ground state degeneracy on a torus, while $\bs P$ has a unique ground state. Therefore, the stabilizer code obtained by replacing $\tbs Z$ with $\tbs H_{M3}$ should also realize different fracton phases. It would be interesting to look into the properties of this model.

\begin{figure*}[t]
        \includegraphics[scale=1]{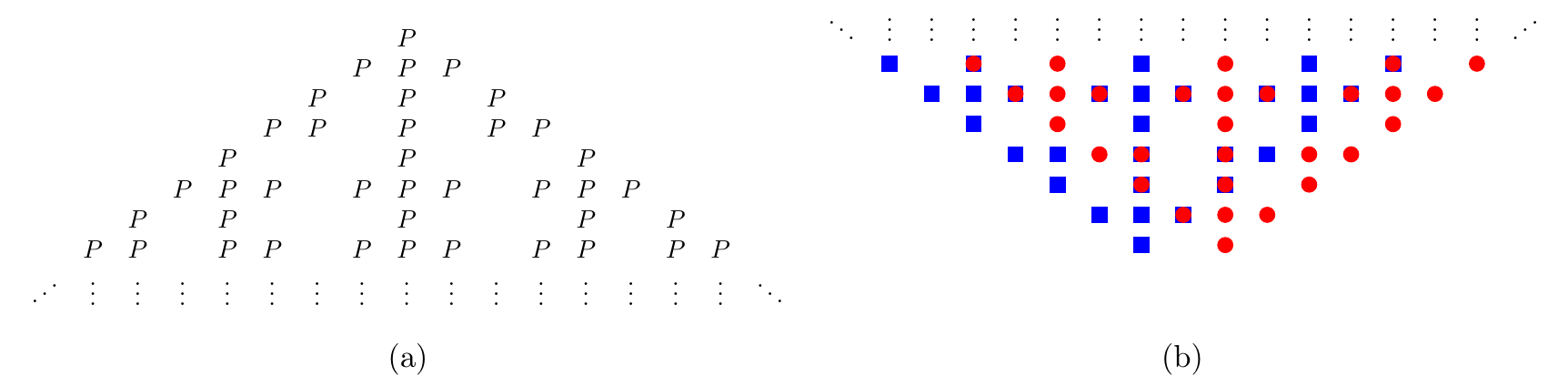}
    \caption{(a) Fractal fermion parity symmetry acting on sites in the pattern of the Fibonacci CA. (b) $\mathbb Z_2$ fractal symmetry in the dual Pauli model. Red circles and blue squares denote Pauli $X$'s and Pauli $Z$'s, respectively. Their overlap denotes Pauli $Y$'s.}
    \label{fig:fractalsym}
\end{figure*}

\subsection{(110) Fermion planar symmetry in 3D}\label{ex:planar3D_2}
Consider the interaction term
\begin{align}
\bs S&= \lc \begin{pmatrix}
x+y+z \\
\hline \bar x +\bar y + \bar z
\end{pmatrix}.
\end{align}
The symmetries of this system are six planar symmetries, given by the $(110)$, $(1\bar 10)$, $(101)$, $(10\bar 1)$, $(011)$, and $(01\bar1)$ planes.

Since, $\inner{\bs S, \bs S}_F =0$, the dual operators are
\begin{align}
\tbs X&= \lc\begin{pmatrix}
0\\
\hline 
1
\end{pmatrix}, &
\tbs Z&= \lc\begin{pmatrix}
x+y+z + \bar x + \bar y+\bar z\\
\hline 
0
\end{pmatrix},
\end{align}
and the dual symmetries are Pauli $X$ operators acting on the six same planes. In this duality, the Majorana checkerboard model\cite{VijayHaahFu2015} whose stabilizer is
\begin{align}
\bs H_{M1}&= \begin{pmatrix}
1+x+y+z\\
\hline1+\bar x + \bar y+\bar z
\end{pmatrix} = \bs \sigma_F \begin{pmatrix}
  1\\1
\end{pmatrix}
\end{align}
dualizes to the 3D cluster state on a cubic lattice
\begin{align}
\tbs H_{M1}&= \tbs \sigma\begin{pmatrix}
  1\\1
\end{pmatrix} = \begin{pmatrix}
x+y+z + \bar x + \bar y+\bar z\\
\hline1
\end{pmatrix}.
\end{align}

\subsection{Fibonacci Fractal symmetry in 2D}\label{ex:Fibonacci2D}
We consider a fermion model with the following interaction term
\begin{equation}
    S= \begin{array}{ccc}
\gamma & \gamma & \gamma\\
&\gamma' &
\end{array},
\end{equation}
which is represented by
\begin{equation}
    \bs S = \lc\begin{pmatrix}
    y(1+x+\bar x)\\
    \hline
    1
    \end{pmatrix}
\end{equation}
in the algebraic notation. The symmetry that protects this phase is generated by fermion parities placed in the fractal shape of the Fibonacci cellular automaton (CA) \begin{equation}
    \ker \bs \epsilon_F  = \lc\begin{pmatrix}
    \sum_i \bar y^i (1+x+\bar x)^i\\
    \hline
    \sum_i \bar y^i (1+x+\bar x)^i
    \end{pmatrix},
\end{equation}
which is depicted visually in Fig. \ref{fig:fractalsym}. Since this CA is reversible, the symmetries are well defined on a torus. 

We calculate the commutation matrix to be $\inner{\bs S,\bs S}_F = x^2 + \bar x ^2$ and so we choose $\bs T= x^2$. The dual operators are
\begin{align}
   \tbs X &=\lc\begin{pmatrix}    
  x^2 \\
    \hline  
    1 \end{pmatrix}, &\tbs Z &=\lc\begin{pmatrix}  1+ \bar y(1+x+\bar x)\\
    \hline
    0
    \end{pmatrix}.
\end{align}
The dual symmetries are also fractal
\begin{equation}
    \ker \tbs \epsilon  = \lc\begin{pmatrix}
    \bar x^2\sum_i  y^i (1+x+\bar x)^i\\
    \hline
    \sum_i  y^i (1+x+\bar x)^i
    \end{pmatrix}
    \label{equ:2Dfractaldualsym},
\end{equation}
and can be thought of as Pauli-$X$ operators applied in the upside-down version of the Fibonacci fractal pattern above, followed by Pauli-$Z$ operators in the same pattern, but displaced two sites in the $-x$ direction, as shown in Fig. \ref{fig:fractalsym}. Visually, it is clear that since the positions of the Pauli $Z$'s are translated to the left, there is no translation invariant circuit that can remove the Pauli $Z$'s. When $\bs Z$ is the stabilizer, it is in the spontaneously broken phase.

\subsection{Fibonacci Fractal symmetry in 3D}\label{ex:Fibonacci3D}
We now consider a similar model in 3D with an extra hopping (i.e., $q=2$ interaction) in the $z$-direction. That is,
\begin{equation}
    \bs S = \lc\begin{pmatrix}
    y(1+x+\bar x) & 1\\
    \hline
    1 & z
    \end{pmatrix}.
\end{equation}
The symmetries are now stacks of the Fibonacci CA in the $z$-direction.
\begin{equation}
    \ker \bs \epsilon_F  = \lc\begin{pmatrix}
    \sum_{ij} \bar y^i (1+x+\bar x)^i z^j\\
    \hline
    \sum_{ij} \bar y^i (1+x+\bar x)^i z^j
    \end{pmatrix}
\end{equation}
The commutation matrix is
\begin{equation}
    \inner{\bs S,\bs S}_F = \begin{pmatrix}
    x^2 + \bar x ^2 & \bar y (1+x+\bar x) +z\\
   y (1+x+\bar x) +\bar z &0
    \end{pmatrix}
\end{equation} and so we choose the transmutation matrix
\begin{equation}
    \bs T = \begin{pmatrix}
    x^2  & \bar y (1+x+\bar x) +z\\
0&0
    \end{pmatrix}.
\end{equation}
The dual operators are
\begin{align}
   \tbs X &=\lc\begin{pmatrix}    
  x^2 & \bar y (1+x+\bar x) +z \\
  0&0\\
    \hline  
    1 & 0\\
    0&1\end{pmatrix}, &\tbs Z &=\lc\begin{pmatrix}  1+ \bar y(1+x+\bar x) \\1+\bar z\\
    \hline
    0\\0
    \end{pmatrix}. \nonumber
\end{align}
In this model, there is a local dual symmetry
\begin{equation}
    \tbs G  = \lc\begin{pmatrix}
    \bar x^2(1+z)\\ 
    (1 + y(1+x+\bar x))(1+\bar z)\\ 
    \hline
    1+z\\
    1+y(1+x+\bar x)\\
    \end{pmatrix}\subset \ker \tbs \epsilon
\end{equation}
This term, along with $\bs Z$, forms a stabilizer code 
\begin{equation}
\lc\begin{pmatrix}
    \bar x^2(1+z) &1+ \bar y(1+x+\bar x)  \\
    (1 + y(1+x+\bar x))(1+\bar z) & 1+\bar z\\
    \hline
    1+z &0\\
    1+y(1+x+\bar x) &0
    \end{pmatrix},
    \label{equ:twistedYoshida}
\end{equation}
which is a twisted version of Yoshida's fractal spin model\cite{Yoshida2013}.

\subsection{Haah's Fractal symmetry}\label{ex:Haah}
In analogy to the fractal Ising model\cite{VijayHaahFu2016,Williamson2016}, consider a fermionic system with interaction terms
\begin{align}
\bs S&= \lc\begin{pmatrix}
f_1 & 0\\
\hline 0 &f_2
\end{pmatrix}
\end{align}
where $f_1 = 1+x+y+z$ and $f_2 =1+xy +yz +zx$. The symmetries of this model are local fermion parities acting on sites in the same fractal pattern as the fractal Ising model\cite{VijayHaahFu2016}. Since $\inner{\bs S,\bs S}_F = \begin{pmatrix} f_1\bar f_1 & 0\\ 0 & f_2 \bar f_2 \end{pmatrix}$, we construct
\begin{equation}
\bs T= \begin{pmatrix}
t_1 & 0\\
0 & t_2 
\end{pmatrix},
\end{equation}
where $t_1 =  x+y+z+x\bar y+y\bar z + z\bar x$ and $t_2= xy+yz+zx+x\bar y+y\bar z + z\bar x$. One can verify that $t_1+\bar t_1 = f_1\bar f_1$ and $t_2+\bar t_2 = f_2\bar f_2$. The dual operators are
\begin{align}
   \tbs X &=\lc
   \begin{pmatrix}t_1 & 0\\
0 & t_2\\
\hline 
1&0\\0&1\end{pmatrix},  &
 \tbs Z &=\lc\begin{pmatrix} \bar f_1 \\ \bar f_2  \\ \hline 0 \\0
 \end{pmatrix}.
\end{align}
The local gauge constraints are generated from
\begin{equation}
\tbs G =
 \lc\begin{pmatrix}
\bar t_1 f_2 \\
\bar t_2 f_1  \\
\hline
f_2\\
f_1\\
\end{pmatrix}.
\end{equation}
Together, $\tbs Z$ and $\tbs G$ are stabilizers that realize a twisted Haah's code. In general, these formulas hold for any model with two types of interaction terms, by replacing $f_1,f_2$ and solving for $t_1,t_2$.

The Majorana Hamiltonian given by $\bs H_{M5} = \lc \begin{pmatrix} f_5 \\ \hline \bar f_5\end{pmatrix}$, where $f_5= 1+x+y+z + xy+yz+xz$ can be generated from fermion parity and the interaction terms. In particular,
\begin{equation}
  \bs H_{M5} = \bs \sigma_F   \lc\begin{pmatrix}1+\bar x \bar y \bar z\\1+\bar x \bar y \bar z\\xy+yz+zx+\bar x \bar y + \bar y \bar z + \bar z \bar x+ \bar x \bar y \bar z \end{pmatrix}.
\end{equation}
The resulting dual Hamiltonian is
\begin{widetext}
\begin{equation}
  \tbs H_{M5} = \begin{pmatrix} 
  x+y+z +xy+yz+zx+x\bar y + y\bar z+ z\bar x +\bar x\bar y^2 + \bar y \bar z^2 + \bar z \bar x^2 + x y \bar z + y z \bar x + z x\bar y +\bar x \bar y \bar z^2 + \bar y \bar z \bar x^2 + \bar z \bar x \bar y^2\\
  1 + xy+yz+zx+ x\bar y + y\bar z+ z\bar x + \bar x\bar y^2 + \bar y \bar z^2 + \bar z \bar x^2 + \bar x^2\bar y^2 + \bar y^2 \bar z^2 + \bar z^2 \bar x^2  + \bar x\bar y\bar z \\
  \hline
1+\bar x \bar y \bar z\\1+\bar x \bar y \bar z\end{pmatrix}
\end{equation}
\end{widetext}
The stabilizer code $(\tbs H_{M5} \ \tbs G)$ should result in a different fracton phase from the stabilizer code $(\tbs Z \ \tbs G)$.
\begin{figure}[h]
    \centering
    \includegraphics[scale=0.23]{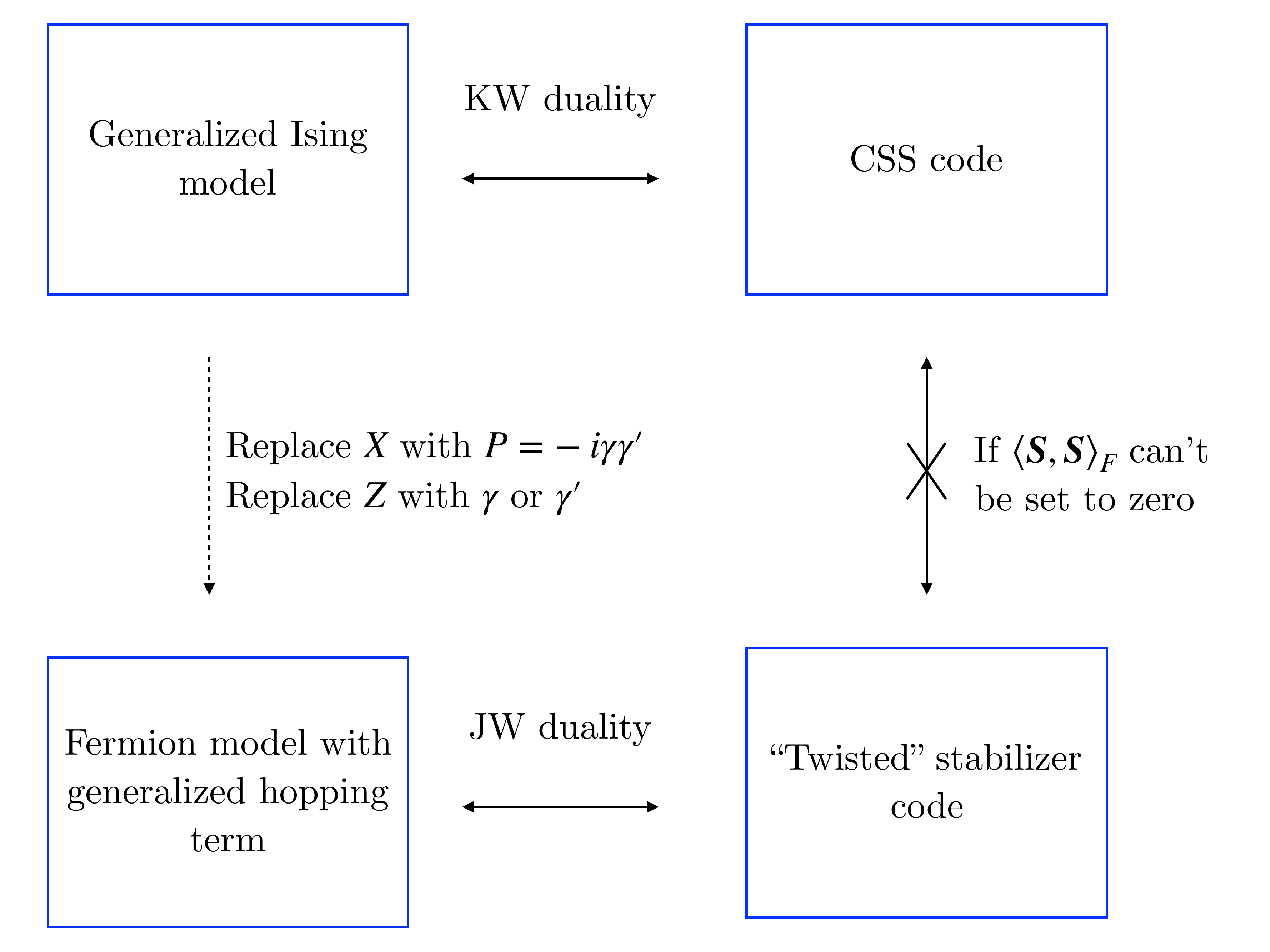}
    \caption{Constructing ``twisted'' stabilizer codes from known CSS codes.}
    \label{fig:twistedconstruction}
\end{figure}
\section{Twisted codes from CSS codes}\label{twistedCSS}
As an application of the KW and JW dualities, we present a general method of constructing twisted codes from CSS codes, which is illustrated in Fig. \ref{fig:twistedconstruction}.
\begin{enumerate}
\item Start from a CSS code and treat the stabilizers containing $X$ terms as the dual symmetry. Add an $X$ transverse field into the Hamiltonian to make it a dual transverse Ising model.
\item Perform a KW duality to obtain an Ising model. This is equivalent to the ``ungauging'' procedure of Ref. \onlinecite{KubicaYoshida2018}.
\item If the Ising terms contain an even number of Pauli $Z$'s, construct a fermion model by replacing transverse fields $X$ with fermion parity $P$ and obtain interaction terms by replacing each Pauli $Z$ with either $\gamma$ or $\gamma'$ (up to a phase $\pm i$).
\item Perform the JW duality on this new fermion model, and construct a new stabilizer model using the Ising terms (which will be identical to the $Z$ stabilizers in the original CSS code), and local terms of the dual symmetry. If there is no choice such that $\inner{\bs S,\bs S}_F=0$ up to attaching local fermion parity operators, then the new stabilizer code cannot be related to the original CSS code via a translation-invariant Clifford unitary.
\end{enumerate}

In fact, many of the examples we have discussed in the previous section are obtained from this procedure. For example, starting from the 3D toric code, one can either ``ungauge'' the 2-form symmetry or 1-form symmetry to get a 0-form or 1-form Ising model, respectively. Replacing this bosonic system with a fermionic system with the same type of symmetries then gives the initial fermion systems for Secs. \ref{ex:Global3D} and \ref{ex:1form}, respectively. We have also used this to construct the twisted versions of X-cube, Yoshida's fractal code, and Haah's code.

We will further demonstrate this procedure with a class of examples. We will choose the CSS code to be a self-dual ``doubled'' Majorana code\cite{BravyiTerhalLeemhuis2010}. As a byproduct, this procedure gives us a natural JW duality to bosonize the Majorana code and produces a new code which is distinct from both the original CSS code and the twisted code.

 The Majorana codes introduced in Ref. \onlinecite{VijayHaahFu2015} are given by a Majorana stabilizer of the form  $\bs H_M =\lc \begin{pmatrix}
f\\
\hline
\bar f
\end{pmatrix}$ for some polynomial $f$. It satisfies $\inner{\bs H_M,\bs H_M}_F=0$. In addition, we will require $f$ to contain an even number of terms.

The doubled CSS code of such a Majorana code is given by the Pauli stabilizer
\begin{equation}
\lc\begin{pmatrix}
\bar f & 0\\
f & 0\\
\hline
0 & \bar f\\
0 &  f
\end{pmatrix}
\end{equation}
which one can verify that all stabilizers commute.

To perform a KW duality, we will treat the first column as the Ising term $\tbs Z$ and the second column as a local symmetry constraint $\tbs G \subset \ker \tbs \epsilon$. The KW dual then gives
\begin{align}
\bs Z &= \lc\begin{pmatrix}
f & \bar f\\
\hline 
0 &0 
\end{pmatrix}, &\bs X &= \lc\begin{pmatrix}
0\\
\hline 1
\end{pmatrix} 
.
\end{align}
The next step is to replace Pauli operators with appropriate Majorana operators.
\begin{align}
\bs S &=\lc \begin{pmatrix}
f & 0 \\
\hline 
0 &\bar f
\end{pmatrix}, &
\bs P &=\lc \begin{pmatrix}
1\\
\hline 1
\end{pmatrix},
\end{align}
where we have replaced the first and second Ising terms with $\gamma$ and $\gamma'$, respectively. The reason of such choice is so that the two types of interaction terms automatically commute. The commutation matrix is
\begin{equation}
 \inner{\bs S,\bs S}_F = \begin{pmatrix}
 f\bar f & 0\\
 0 & f\bar f
 \end{pmatrix}.  
\end{equation}
From Lemma \ref{prop:Texists}, a transmutation matrix $\bs T$ exists. This implies that there is a polynomial $t$ such that $t+\bar t = f\bar f$. Therefore, we can choose
\begin{equation}
 \bs T= \begin{pmatrix}
t & 0\\
 0 &\bar t
 \end{pmatrix},  
\end{equation}
and the dual operators are given by
\begin{align}
\tbs X' &= \lc\begin{pmatrix}
t & 0\\
 0 &\bar t\\
\hline 1 &0\\
0&1
\end{pmatrix}, & 
\tbs Z &= \lc\begin{pmatrix}
\bar f\\
f\\
\hline 
0 \\0 
\end{pmatrix}.
\end{align}
Note that here, $\tbs Z$ remains the same. The local gauge constraints are obtained from the identity generator
\begin{align}
    g= \begin{pmatrix}
      \bar f\\
      f\\
      f\bar f
    \end{pmatrix} \in \ker \bs \sigma_F
    \end{align}
This gives
\begin{align}
\tbs G' =\tbs \sigma g= \lc\begin{pmatrix}
\bar t \bar f\\
t f\\
\hline
\bar f\\
f
\end{pmatrix}.
\end{align}
To conclude, the twisted stabilizer code is given by
\begin{align}
\lc \begin{pmatrix}
\bar f &\bar t\bar f\\
f &t f\\
\hline
0&\bar f\\
0& f
\end{pmatrix}.
\end{align}
In addition, the JW duality can also bosonize the original Majorana code. This is because $\bs H = \bs S_1 + \bs S_2$. Hence, its dual is given by
\begin{align}
\tbs H &=\lc \begin{pmatrix}
t\\
\bar t\\
\hline
1\\
1
\end{pmatrix},
\end{align}
and the new stabilizer code from gauging the original Majorana model is
\begin{align}\lc
 \begin{pmatrix}
t &\bar t\bar f\\
\bar t &t f\\
\hline
1&\bar f\\
1& f
\end{pmatrix}.
\label{equ:gaugedMajorana}
\end{align}
An example of this calculation is to use $f_1=1+x+y+z$. The duality can be obtained by choosing $t=x+y+z+ x\bar y + y\bar z+ z\bar x$. The symmetries of the fermion model are planar symmetries on an FCC lattice \cite{VijayHaahFu2016}. The stabilizer given by Eq. \eqref{equ:gaugedMajorana} is the result of gauging the Majorana checkerboard model with such planar symmetries and should realize a different fracton phase from its double, the Pauli checkerboard model.

One can also do the same calculation for the Majorana model given by $f_4 =1+y+z+xy+yz+xz$, by choosing $t=xy+yz+zx +x (\bar y+ \bar z)+ (x+\bar x)y\bar z$.

\section{Emergent Fermions and Anomalies}\label{'tHooft}
The JW transformation developed allows us to construct spin models which seem to have emergent fermions. However, defining what it means for such a particle to be a ``fermion''  seems to be a subtle issue. A fermion is usually defined via its exchange statistics, which can be computed via carefully designed braiding processes in the lattice model\cite{LevinWen2003}. However, in many of the cases we have considered (specifically where the emergent particles are also fractons), it is not clear how to exchange such particles if they are also immobile.

An alternative way that has been used to imply the existence of fermions is the existence of an anomaly. One method is to argue that the symmetry cannot be consistently coupled to a dynamical gauge field and so has an 't Hooft anomaly \cite{GaiottoKapustin2016}. This physically corresponds to the inability to condense the fermion. Alternatively, one can argue at the lattice model level that the symmetry cannot be realized in an onsite manner, and correspondences between the two have been established\cite{Wen2019}.

However, a field theory description for models with such exotic symmetries is still in development and it is not obvious how to properly define support for such symmetries. For example for fractal symmetries, sites outside the support of the original fractal could alternatively be considered inside the support of a different fractal. Unfortunately, rigorously arguing whether a subsystem or higher-rank symmetries has an anomaly by the above methods is beyond the scope of this work.

In this paper, the best we are able to argue that might be indicative of an anomalous $\mathbb Z_2$ symmetry is to argue that a nonanomalous symmetry is one that can be written in a form that consists of only a single type of Pauli matrix (in this case, Pauli-$X$). This makes the symmetry factorizable into a tensor-product structure, allowing a KW duality, which suggests that it can be coupled to a dynamical gauge field. On the other hand, an obstruction to having such form is indicative of some anticommutations that results from having additional Pauli $Z$'s in the symmetries that cannot be removed. Furthermore, such a symmetry does not admit a KW dual in our formalism, and so in some sense can be related to the inability to condense the excitations.

If we assume this criteria as a partial indication of the anomaly, we are able to show the following:
\begin{prop}
The spin system JW dual to the fermion system as defined cannot have a corresponding KW dual unless $\inner{\bs S,\bs S}_F=0$.
\begin{proof}
The JW dual to the fermion system has dual symmetry
\begin{equation}
    \ker \tbs \epsilon = \ker  \begin{pmatrix}[c!{\color{\linecolor}\vrule} c] \mathbbm 1 & \bs T^\dagger\\
  0 &\inner{\boldsymbol P,\boldsymbol S}_F \end{pmatrix}.
\end{equation}
Let us assume such a KW dual exists, i.e., there exists $\bs \sigma = (\bs X \ \bs Z)$ such that the KW diagram \eqref{equ:sequence2} commutes. Then, up to a basis transformation we must also have
\begin{equation}
    \ker \tbs \epsilon =  \ker \begin{pmatrix}[c!{\color{\linecolor}\vrule} c] \mathbbm 1 & 0\\
  0 &\inner{\boldsymbol X,\boldsymbol Z} \end{pmatrix},
\end{equation}
which contains symmetries that have only Pauli $X$'s. For this to be satisfied, there must exist a symplectic matrix $\bs U = \begin{pmatrix} \bs U_1 &\bs U_2\\
  \bs U_3 &\bs U_4 \end{pmatrix}$ such that
\begin{equation}
    \begin{pmatrix}[c!{\color{\linecolor}\vrule} c] \mathbbm 1 & 0\\
  0 &\inner{\boldsymbol X,\boldsymbol Z} \end{pmatrix} \bs U  =
 \begin{pmatrix}[c!{\color{\linecolor}\vrule} c] \mathbbm 1 & \bs T^\dagger\\
  0 &\inner{\boldsymbol P,\boldsymbol S}_F \end{pmatrix}.
\end{equation}
Solving this gives $\bs U_1=\mathbbm 1$, $\bs U_2 = \bs T^\dagger$, $\bs U_3 = 0$, $\inner{\boldsymbol X,\boldsymbol Z} \bs U_4 = \inner{\boldsymbol P,\boldsymbol S}$. Imposing $\bs U$ is symplectic i.e., $\bs U^\dagger \bs \lambda \bs U = \bs \lambda$, gives $\bs U_4 = \mathbbm 1$ and $\bs T + \bs T^\dagger = \inner{\bs S,\bs S}_F =0$.
\end{proof}
\end{prop}
Therefore, we must show that it is impossible to choose the interaction terms in the fermion system such that they all commute. In Sec. \ref{Examples}, we have explicitly written down various models where $\inner{\bs S,\bs S} \ne 0$. Hence, we need to justify that one cannot redefine commuting interaction terms simply by attaching local fermion parities. From Prop \ref{prop:dualsymmetryinvariant}, this implies that there are no polynomials $f_{ki} \in R$ such that
\begin{equation}
\inner{\bs S_i,\bs S_j}_F +\inner{\bs S_i,\bs P_k}_Ff_{kj} + \inner{\bs P_k,\bs S_j}_F\bar f_{ki}  =0.
\label{eq:anomalyfree}
\end{equation}
For the examples in this paper, we are able to prove this for the 2D and 3D twisted toric codes, and the 2D and 3D models with Fibonacci fractal symmetry. A proof (assuming translation invariance) can be found in Appendix \ref{app:anomalyproof}. This implies that the dual symmetries of the Fibonacci fractal models in Secs. \ref{ex:Fibonacci2D} and \ref{ex:Fibonacci3D} are indeed anomalous. We relegate the task of determining whether the other models we have constructed are anomalous or not to future work.

\subsection{Anomaly cancellation example}
Due to the bulk-boundary correspondence, theories with anomalies can be canceled by an SPT in one higher dimension. For example, the 1-form anomaly in 2D can be canceled by a 3D 1-form SPT\cite{KapustinThorngren2017,TsuiWen2020}. Furthermore, the SPT itself is dual to the ground state of a 3D twisted toric code, and so at low energies, it has an emergent anomalous 2-form symmetry, which can be canceled by a 4+1D bulk of a 2-form SPT. The hierarchy of anomalies and SPTs continues in this fashion ad infinitum\cite{ChenPC}.

Here, we will demonstrate an example with a similar hierarchy. The JW dual of the 2D Fibonacci fermion model in Sec. \ref{ex:Fibonacci2D} lives naturally on the boundary of a 3D SPT with fractal symmetry. This SPT can then be KW dual to a model whose ground state realizes the twisted fractal spin model in Sec. \ref{ex:Fibonacci3D}.

First, let us consider the 3D model with the Fibonacci Ising term in the $xy$ plane and a standard Ising term in the $z$ direction, reminiscent of the fermion model in Sec. \ref{ex:Fibonacci3D}:
\begin{align}
    \bs Z &=\lc \begin{pmatrix}
    1+y(1+x+\bar x) & 1+z\\
    \hline
    0 &0
    \end{pmatrix}, & \bs X = \lc\begin{pmatrix}
   0\\ \hline 1
    \end{pmatrix}.
\end{align}
The symmetries are stacks of the Fibonacci CA in the $z$ direction.
\begin{equation}
    \ker \bs \epsilon =\lc \begin{pmatrix}
    0\\
    \hline
   \sum_{ij}\bar y^i(1+x+\bar x)^iz^j
    \end{pmatrix}.
\end{equation}
The KW dual of this model is given by
\begin{align}
   \tbs X &=\lc\begin{pmatrix}    
  0 & 0 \\
  0&0\\
    \hline  
    1 & 0\\
    0&1\end{pmatrix}, &\tbs Z &=\lc\begin{pmatrix}  1+ \bar y(1+x+\bar x) \\1+\bar z\\
    \hline
    0\\0
    \end{pmatrix},
\end{align}
with dual symmetry constraints 
\begin{equation}
    \ker \tbs \epsilon  = \lc\begin{pmatrix}
    0 &0& 0\\
   0 &0 &0\\
    \hline
    1+z&\sum_i  y^i (1+x+\bar x)^i&0\\
    1+y(1+x+\bar x)&0 & \sum_i z^i
    \end{pmatrix}.
\end{equation}
The first column of $\ker \tbs \epsilon$ and $\tbs Z$ commute. Therefore, together, they form the stabilizer code
\begin{equation}
\lc\begin{pmatrix}
    0&1+ \bar y(1+x+\bar x)  \\
   0 & 1+\bar z\\
    \hline
    1+z &0\\
    1+y(1+x+\bar x) &0
    \end{pmatrix}
    \label{equ:untwistedYoshida}
\end{equation}
for the untwisted fractal spin model.
Now, consider the following stabilizer:
\begin{align}
    \bs H_{SPT}&= \bs \sigma \begin{pmatrix}
   1+z+(1+\bar z)\bar y(1+x+\bar x)\\
   x^2(1+\bar z) \\
    1
    \end{pmatrix} = \lc\begin{pmatrix}
     A_{SPT}\\
     \hline
   1
    \end{pmatrix},
    \label{equ:fractalSPT}
\end{align}
where $A_{SPT}= x^2 + \bar x^2 +(y+ \bar y) (1+x+\bar x) +z (1+ x^2 +y(1+x+\bar x))+ \bar z(1+\bar x^2 + \bar y(1+x+\bar x))$. One can check that $\inner{\bs H_{SPT},\bs H_{SPT}}=0$. Furthermore, since it corresponds to a generator label, it respects the symmetry, and can be dualized to
\begin{align}
    \tbs H_{SPT}= \tbs \sigma \begin{pmatrix}
   1+z+(1+\bar z)\bar y(1+x+\bar x)\\
   x^2(1+\bar z)\\
    1
    \end{pmatrix}\nonumber\\
    =\lc \begin{pmatrix}
    1+\bar y(1+x+\bar x)\\
    1+\bar z\\
    \hline
    (1 + \bar y(1+x+\bar x))(1+ z)\\
    x^2(1+\bar z)
    \end{pmatrix}.
\end{align}
The first column of $\ker \tbs \epsilon$ and $\tbs H_{SPT}$ are together the stabilizer code
\begin{equation}
\lc\begin{pmatrix}
    0&1+ \bar y(1+x+\bar x)  \\
   0 & 1+\bar z\\
    \hline
    1+z &(1 + \bar y(1+x+\bar x))(1+ z)\\
    1+y(1+x+\bar x) & x^2(1+\bar z)
    \end{pmatrix},
\end{equation}
which is exactly the twisted fractal spin model \eqref{equ:twistedYoshida} up to inversion and an appropriate swap of rows and columns. When $\tbs H_{SPT}$ is enforced energetically, the low-energy Hilbert space has an emergent symmetry (given by the stabilizers of $\tbs H_{SPT}$) which is anomalous.

Since the twisted and untwisted fractal spin models cannot be connected via a translation-invariant Clifford circuit, we have confirmed that $\bs H_{SPT}$ in Eq. \eqref{equ:fractalSPT} is an SPT protected by the fractal symmetry. In particular, since it is a cluster state, when the symmetry is explicitly broken, the model can be disentangled with controlled-$Z$ gates corresponding to a symplectic transformation $\begin{pmatrix} 1 & A_{SPT}\\ 0 &1\end{pmatrix}$.

Let us now look at the symmetry action on the boundary of $\bs H_{SPT}$. We consider a semi-infinite 3D system from $z= -\infty$ which terminates at $z=0$. The symmetry is given by
\begin{equation}
  \lc\begin{pmatrix}
    0\\
    \hline 
   \sum_{i}\sum_{j=-\infty}^0 \bar y^i(1+x+\bar x)^i z^j
    \end{pmatrix}.
\end{equation}
In the ground state, we can use $\bs H_{SPT}$ to substitute a Pauli $X$ with Pauli $Z$'s at the positions given by the first row of $\bs H_{SPT}$. However, since $\bs H_{SPT}$ spans three layers in the $z$ direction, this substitution is only valid from the second layer downwards. A calculation in Appendix \ref{app:symmetryaction} shows that the effective symmetry action at the layer $z^0$ matches the anomalous symmetry action of the JW dual with Fibonacci symmetry \eqref{equ:2Dfractaldualsym} up to inversion. This confirms that $\bs H_{SPT}$ is indeed an SPT.

\section{Discussion}\label{Discussion}
Assuming translation invariance, we have constructed a generalization of the JW duality that performs an exact bosonization of a fermion system with arbitrary $q$-body interactions. Under this framework, we have proven the existence and uniqueness (up to a choice of basis) of the dual spin theory.

In the case of multibody interaction terms, the fermionic Hamiltonian has an additional higher-form or subsystem fermion parity symmetry and the dual spin theories can in some cases exhibit fracton topological order. Furthermore, starting from a CSS code, the dualities allow us to construct a new ``twisted'' stabilizer code with possible fermionic excitations, and at the same time bosonize Majorana codes in various ways.

We conclude by listing many open questions.
\begin{enumerate}
\item Fermionic nature of fractons: From the duality, exactly solvable models (stabilizer codes) can be constructed by properly ``gauging'' the $\mathbb Z_2^F$ global/higher form/subsystem symmetries. In the case of planar or fractal symmetries in 3D, the resulting models are twisted models of the X-cube, checkerboard, Haah's code, and Yoshida's fractal code. The fracton excitations are fermionic in the sense that there is an obstruction to condensing them. It would be interesting to see if there is a meaningful exchange procedure to detect the ``fermionic statistics'' of these fractons. We intend to address this in future work. Furthermore, it would be interesting to show whether these twisted models are in different phases from their usual counterparts, either in the usual sense of a quantum phase, possibly including translation symmetries\cite{PaiHermele2019} or in terms of foliated fracton order\cite{ShirleySlagleChen2019,DevakulShirleyWang2019}.

\item 't Hooft anomalies: If the fractons truly have a fermionic nature, then there should be an associated 't Hooft anomaly which generalizes the Steenrod square topological action.\cite{GaiottoKapustin2016,ZhuLanWen2019,LanZhuWen2019,TsuiWen2020, Chen2019}. As field theoretical methods are being developed to describe fracton phases\cite{SlagleKim2017,Seiberg2019,SeibergShao2020_1,SeibergShao2020_2,SeibergShao2020_3}, do our proposed lattice models, when transcribed into the field theory language, have the correct anomaly? A closely related question is due to the bulk-boundary correspondence: What is the corresponding SPT in one higher dimension that has the corresponding anomaly on its boundary?

\item Given a certain spin model, is there a way to check if it is dual to some fermionic system? For example, for models whose excitations are known to have fermion statistics, such as the Levin-Wen fermion model\cite{LevinWen2003}, or spin models which admit Parton constructions in terms of Majorana fermions\cite{Kitaev2006,Ryu2009,NussinovOrtizCobanera2012,HsiehHalasz2017}, how does one determine the associated symmetries and JW dual of these models? 

\item General lattices: Although translation invariance was a key assumption in establishing the dualities in this paper, the existence of dual Pauli operators with the same commutation relations in arbitrary lattices can be similarly argued to exist by replacing the commutation matrix $\inner{\bs S,\bs S}_F$ with a large ``adjacency matrix'' between all interaction terms that anticommute in an arbitrary lattice, and constructing the analog of the transmutation matrix $\bs T$ as an upper triangular matrix. Nevertheless, a necessary condition for a self-consistent duality in the global symmetry case is the vanishing of the second Stiefel-Whitney of the manifold, and the duality depends on a choice of spin structure\cite{GaiottoKapustin2016,TarantinoFidkowski2016,Wareetal2016,EllisonFidkowski2019,TantivasadakarnVishwanath2018}. For general interaction terms, are there similar obstructions that generalize the notion of Stiefel-Whitney classes and spin structures? For example, the duality between the fermion model with planar symmetry in Sec. \ref{ex:planar3D_1} and the twisted X-cube model -- if well defined on a 3-torus -- would actually depend on a choice of $2^{6L-3}$ such ``spin structures''. An interesting extension would be to determine the such obstructions in dualities for fracton models with arbitrary foliations \cite{ShirleySlagleWangChen2018,TianSampertonWang2020} or even perhaps arbitrary cellulations\cite{Radicevic2019}.

\item Fermionic higher-form/subsystem SPTs:  So far, the Hamiltonians on the fermionic side have been either symmetry-breaking or topological ordered. Are there examples of SPTs protected by higher form or subsystem symmetries, i.e., those that are nontrivial solely by subdimensional fermion parity without any further symmetries? Can higher-form fermionic phases be classified by a variant of spin cobordism\cite{Kapustin2014,Kapustinetal2015}?

\item Parafermions: Generalizations to dualities between parafermions and $\mathbb Z_p$ clock models\cite{FradkinKadanoff1980,Radicevic2018} for prime $p$ are possible in this formalism by instead working with the polynomial ring over $\mathbb F_p$. However, in 3D, mobile particles cannot have emergent parafermionic statistics. If fermionic statistics can be properly defined for fractons, can they still be extended to parafermions?

\item It seems that the twisted X-cube model constructed can be obtained from a recent defect network construction in Ref. \onlinecite{Aasenetal2020} by replacing the 3D toric code with the twisted 3D toric code. It would be interesting to see whether such defect construction can account in general for the twisted models we have presented here in a similar fashion.

\item Quantum codes and simulations of  fermions: the dualities presented construct various new stabilizer codes which can be useful for quantum computation. In fact, we have demonstrated multiple ways to bosonize a given fermionic Hamiltonian (depending on the symmetry we chose to ``gauge''). It would be interesting to see which of these bosonized codes are most efficient, for example by looking at the code distance or the average cost for each logical operation.
\end{enumerate}
\textbf{Note Added:} Recently, I became aware of related work by Wilbur Shirley\cite{Shirley2020}, which constructs similar fracton models that have immobile fermion excitations. Our results were obtained independently.

\begin{acknowledgments}
I would like to thank Xie Chen, Yu-An Chen, Tyler Ellison, Jeongwan Haah, Michael Hermele, Sheng-Jie Huang, \DJ or\dj e Radi\v cevi\'c, Thomas Schuster, Wilbur Shirley, Hao Song, Sagar Vijay, Ashvin Vishwanath, and Juven Wang for stimulating discussions. In particular, I am grateful to \DJ or\dj e Radi\v cevi\'c for his patience in explaining to me his work, Jeongwan Haah for explaining how to solve Laurent polynomial equations using the fraction field, and Yu-An Chen, Tyler Ellison, Thomas Schuster, and Sagar Vijay for collaborations on related works and innumerable discussions. I would also like to acknowledge helpful conversations with the participants of the Simons Collaboration on Ultra Quantum Matter Workshop, which was supported by a grant from the Simons Foundation (651440), and the participants of the ``Fractons and Beyond'' workshop at the Banff International Research Station (20w5064). I acknowledge the support of NSERC.
\end{acknowledgments}

\appendix

\section{Proof of correspondence between symmetries and identity generators}\label{app:KWJWconstraints}

In this appendix, we provide proofs giving a one-to-one correspondence between symmetries and identity genertors for the KW dualities (Props. \ref{prop:KWconstraint1} and  \ref{prop:KWconstraint2}) and the JW dualities (Props. \ref{prop:JWconstraint1} and  \ref{prop:JWconstraint2}). 

To prove the equalities, we will need to prove inclusion in both directions. The inclusion to the left can be proven assuming only the commutativity of the diagrams \eqref{equ:sequence2} and \eqref{equ:sequence3}. Let us demonstrate this for Prop. \ref{prop:KWconstraint1}. The proofs for the remaining claims are identical.
\begin{lemma}  In the KW duality, $  \ker \bs \epsilon \supseteq \bs \sigma \ker \tbs \sigma$ 
\begin{proof}
 Let $a \in \ker \tbs \sigma$. It suffices to show that $\bs \epsilon (\bs \sigma a)=0$. Indeed, since the diagram commutes,  $(\bs \epsilon \circ \bs \sigma) a = (\tbs \epsilon \circ \tbs \sigma) a=0$.
 \end{proof}
 \end{lemma}
 
 Now we will prove the inclusion to the right. Starting with the KW duality, we will assume that the Ising term has the form $\bs Z =\lc \begin{pmatrix}
   \bs Z_Z\\ \hline \bs Z_X
 \end{pmatrix}$.
 The generating and excitation maps in this case are
 \begin{align}
   \bs  \sigma &= \lc\begin{pmatrix}
   \bs Z_Z & 0\\ \hline \bs Z_X & \mathbbm 1
 \end{pmatrix}, &\tbs  \sigma &= \lc\begin{pmatrix}  0& \bs Z_Z^\dagger \\\hline \mathbbm 1 & 0\end{pmatrix},\\
   \bs \epsilon &=\begin{pmatrix}[c!{\color{\linecolor}\vrule} c] \bs Z_X^\dagger & \bs Z_Z^\dagger\\ \mathbbm 1 &0 \end{pmatrix}, & \tbs \epsilon &=\begin{pmatrix}[c!{\color{\linecolor}\vrule} c] \mathbbm 1 & 0\\
  0 &\bs Z_Z \end{pmatrix}.
\end{align}
 
\begin{lemma} In the KW duality, $  \ker \bs \epsilon \subseteq \bs \sigma \ker \tbs \sigma$.
\begin{proof}
Let $b=\lc\begin{pmatrix} b_Z\\ \hline  b_X \end{pmatrix} \in \ker \bs \epsilon$, then we must have 
\begin{align}
   b_Z=\bs Z_Z^\dagger b_X=0. 
\end{align}
We want to show that there exists $a \in \ker \tbs \sigma$ such that $\bs \sigma a =b$. Indeed, let $a=\begin{pmatrix} 0\\  b_X \end{pmatrix}$.  Then  $\tbs \sigma a  =0$ and $\bs \sigma a  = b$ as desired.
\end{proof}
\end{lemma}

\begin{lemma}
 In the KW duality, $\ker \tbs \epsilon \subseteq \tbs \sigma \ker \bs \sigma$.
 \begin{proof}
 Let $b=\lc\begin{pmatrix} b_Z\\ \hline  b_X \end{pmatrix} \in \ker \tbs \epsilon$, then we must have
  \begin{align}
      b_Z = \bs Z_Z b_X=0.
  \end{align}
  Let $a=\begin{pmatrix} b_X\\  \bs Z_Xb_X \end{pmatrix}$, then one can check that $a \in \ker \bs \sigma$ since $\bs \sigma a=0$. Furthermore, using $\inner{\bs Z,\bs Z}=\bs Z_Z^\dagger \bs Z_X+ \bs Z_X^\dagger \bs Z_Z=0$,
 \begin{align}
     \tbs \sigma a& =\lc\begin{pmatrix}0 &\bs Z_Z^\dagger\\\hline  \mathbbm 1 & 0 \end{pmatrix} \begin{pmatrix} b_X\\ \bs Z_Xb_X \end{pmatrix}  =\lc\begin{pmatrix} \bs Z_Z^\dagger\bs Z_X b_X \\ \hline b_X\end{pmatrix} \nonumber\\
    & =\lc\begin{pmatrix} \bs Z_X^\dagger\bs Z_Z b_X \\\hline b_X\end{pmatrix} =\lc\begin{pmatrix} 0 \\\hline b_X\end{pmatrix}  =b.
 \end{align}
 \end{proof}
\end{lemma}

The proofs for the JW dualities are nearly identical. First, we assume the interaction terms are of the form
$\bs S =\lc \begin{pmatrix}
   \bs S_Z\\ \hline \bs S_X
 \end{pmatrix}$.
 The generating and excitation maps are then given by
 \begin{align}
   \bs  \sigma_F &= \lc\begin{pmatrix}
   \bs S_Z & \mathbbm 1\\ \hline \bs S_X & \mathbbm 1
 \end{pmatrix}, &\tbs  \sigma &= \lc\begin{pmatrix}\bs T &\bs S_Z^\dagger+\bs S_X^\dagger\\\hline  \mathbbm 1 & 0 \end{pmatrix},\\
   \bs \epsilon_F &=\begin{pmatrix}[c!{\color{\linecolor}\vrule} c] \bs S_Z^\dagger & \bs S_X^\dagger\\ \mathbbm 1 &\mathbbm 1 \end{pmatrix}, & \tbs \epsilon &=\begin{pmatrix}[c!{\color{\linecolor}\vrule} c] \mathbbm 1 & \bs T^\dagger\\
  0 &\bs S_Z+ \bs S_X \end{pmatrix}.
\end{align}

\begin{lemma} In the JW duality, $  \ker \bs \epsilon_F \subseteq \bs \sigma_F \ker \tbs \sigma$.
\begin{proof}
Let $b=\lc\begin{pmatrix} b_Z\\ \hline  b_X \end{pmatrix} \in \ker \bs \epsilon_F$, then we must have 
\begin{align}
   b_Z+b_X =(\bs S_Z + \bs S_X^\dagger) b_Z=0.
\end{align}
Let $a=\begin{pmatrix} 0\\  b_Z\end{pmatrix}$, then one verifies that $\tbs \sigma a   =0$ and $\bs \sigma_F a = b$ as desired.
\end{proof}
\end{lemma}

\begin{lemma}
 In the JW duality, $ \ker \tbs \epsilon \subseteq  \tbs \sigma \ker \bs \sigma_F$.
 \begin{proof}
 Let $b=\lc \begin{pmatrix} b_Z\\ \hline  b_X \end{pmatrix} \in \ker \tbs \epsilon$, then we must have
  \begin{align}
      b_Z+\bs T^\dagger b_X=(\bs S_Z+ \bs S_X)b_X=0
  \end{align}
Let $a=\begin{pmatrix} b_X\\  \bs S_Xb_X \end{pmatrix}$, then $a \in \ker \bs \sigma_F$ since $\bs \sigma_F a  =0$. Furthermore, using $\bs T+ \bs T^\dagger = \inner{\bs S, \bs S}_F = \bs S_Z^\dagger \bs S_Z+\bs S_X^\dagger \bs S_X$,
 \begin{align}
     \tbs \sigma a &=\lc\begin{pmatrix}\bs T &\bs S_Z^\dagger+\bs S_X^\dagger\\\hline  \mathbbm 1 & 0 \end{pmatrix} \begin{pmatrix} b_X\\  \bs S_Xb_X \end{pmatrix}\nonumber\\
     &=\lc\begin{pmatrix} (\bs T + \bs S_X^\dagger\bs S_X + \bs S_Z^\dagger\bs S_X)  b_X \\\hline b_X\end{pmatrix} \nonumber \\
     & =\lc\begin{pmatrix}(\bs T^\dagger+ \bs S_Z^\dagger ( \bs S_Z+ \bs S_X)) b_X \\\hline b_X\end{pmatrix}  =b.
 \end{align}
 \end{proof}
\end{lemma}

\section{Dual of Majorana codes in 3D with global symmetry}\label{app:Majoranaduals}
The models in Ref. \onlinecite{VijayHaahFu2015} are defined as
\begin{equation}
    \bs H_M^{(i)} \equiv \lc\begin{pmatrix} \bar f_i \\ \hline f_i \end{pmatrix}.
\end{equation}
For some polynomial $f_i \in \mathbb F_2[x^{\pm1},y^{\pm1},z^{\pm1}]$. To bosonize, we need to write the Hamiltonian in terms of $\bs P$ and $\bs S$ given by Eq. \eqref{equ:S3D}. That is, we need to find generator labels $\bs g \in G$ such that $\bs \sigma_F \bs g = \bs H_M^{(i)}$. Doing so, we can then obtain the bosonized operator as $\tbs H_M^{(i)} =\tbs \sigma \bs g$. The results are summarized in Table \ref{tab:JW3Dsummary}. Note that the dual operators are only unique up to multiplying by the local symmetry constraints, which are set to one by the duality.

We remark that since the Majorana checkerboard model ($\bs H_M^{(1)}$) is unitary equivalent to a product of the Semionic X-cube model with trivial fermions\cite{WangShirleyChen2019}, the model given by $\tbs H_M^{(1)}$ and local constraints from Eq. \eqref{equ:2formF} should be unitary equivalent to a product of the semionic X-cube model and the twisted 3D toric code.

\begin{table*}[t]
\caption{Duality of Majorana Hamiltonians $\bs H_M^{(i)} \in M$ introduced in Ref. \onlinecite{VijayHaahFu2015} under global fermion parity to its Pauli dual $\tbs H_M^{(i)} \in \tilde P$ with 2-form symmetry constraint given by Eq. \eqref{equ:2formF}. Here, $\bs g_i$ is a generator label such that $\bs \sigma_F \bs g_i = \bs H_F^{(i)}$ and the dual Hamiltonian is given by $\tbs H_M^{(i)} = \tbs \sigma \bs g_i$.}
    \centering
    \begin{tabular}{|c|c|c|}
    \hline
    $f_i$ & $\bs g_i$ & $\tbs H_M^{(i)}$\\
    \hline
    $f_1 =1+x+y+z$ &  $    \begin{pmatrix} 1+\bar x \\ 1+\bar y \\ 1+\bar z \\ 1 \end{pmatrix}$ & 
    $\lc\begin{pmatrix}
    1+\bar x + \bar z + \bar x y\\
     1+\bar y + \bar x + \bar y z\\
       1+\bar z + \bar y + \bar z x\\ \hline  1+\bar x \\ 1+\bar y\\1+\bar z \end{pmatrix}$\\
    \hline
      $f_2 =1+z+xy+yz+xz$ &   $    \begin{pmatrix} y+z+\bar x( \bar y +  \bar z) \\
      0 \\
      y+\bar y \bar z \\
      1+z+\bar z \end{pmatrix}$ & $\lc\begin{pmatrix} 
      1 + \bar x + y + z  + \bar z + \bar x \bar z + \bar y \bar z + \bar x z\\
      1 +y + z +\bar y+ \bar z + \bar y^2 + \bar x \bar y  +\bar x \bar z + \bar y \bar z + \bar y z \\
      x +z + \bar y \bar z + x y \bar z 
      \\ \hline 
      y+ z +\bar x(\bar y+\bar z)\\
      0\\
      y+\bar y \bar z
      \end{pmatrix}$\\
    \hline
      $f_3 =1+x+y+yz+xz$ &  $\begin{pmatrix} 
      1 + \bar x + z + \bar x\bar z \\
      1 + \bar y + z + \bar y\bar z  \\
      0 \\
      1
      \end{pmatrix}$ & $\lc\begin{pmatrix} 
      1 + \bar x y + \bar x \bar z + \bar x y z \\
      \bar x + \bar y + z + \bar x \bar z \\
      x +z+ \bar y + \bar z^2 + x\bar z + \bar y \bar z)  \\
      \hline
     1 + \bar x + z + \bar x\bar z \\
      1 + \bar y + z + \bar y\bar z \\
      0 
      \end{pmatrix}$\\
    \hline
      $f_4 =1+y+z+xy+yz+xz$ &  $\begin{pmatrix} y+z+\bar x( \bar y +  \bar z) \\ 1+\bar y \\y+\bar y \bar z \\ z+\bar z \end{pmatrix}$ & $\lc\begin{pmatrix} 
       \bar x + y + z  + \bar z + \bar x \bar z + \bar y \bar z + \bar x z+\bar x y\\
     y + z + \bar z + \bar y^2 + \bar x \bar y  +\bar x \bar z + \bar y \bar z + \bar y z \\
      x +\bar y+\bar z +z \bar y \bar z + x y \bar z
      \\ \hline 
       y+ z +\bar x(\bar y+\bar z)\\
      1+\bar y\\
      y+\bar y \bar z
      \end{pmatrix}$\\
    \hline
      $f_5 =1+x+y+z+xy+yz+xz$ &  $    \begin{pmatrix} 
      y+\bar x \bar y \\
      z+\bar y \bar z \\
      x +\bar z \bar x \\
     1+x+\bar x + y + \bar y + z + \bar z
      \end{pmatrix}$ & $\lc\begin{pmatrix} 
      y+ \bar y  + z + \bar z + \bar x(\bar x + y  + \bar y  +  z + \bar  z + y z) \\
     z+ \bar z  + x + \bar x + \bar y(\bar y + z  + \bar z  +  x + \bar  x + z x) \\
     x+ \bar x  + y + \bar y + \bar z(\bar z + x  + \bar x  +  y + \bar  y + x y) \\
      \hline
     y+\bar x \bar y \\
     z+\bar y \bar z \\
      x +\bar z \bar x 
      \end{pmatrix}$\\
    \hline
      $f_6 =1+x+y+z+yz$ &  $    \begin{pmatrix} 
      1+\bar x \\
      0 \\
      1+ y+\bar z + \bar y \bar z \\
     1+y + \bar y
      \end{pmatrix}$ & $\lc\begin{pmatrix} 
      \bar x + \bar y +\bar z + \bar x \bar y + \bar x y  + \bar y \bar z, \\
     1 + \bar x + \bar y + y + z + \bar y z \\
     1 +  y+ \bar y  + \bar z x + \bar y \bar z +\bar z y \\
      \hline
    1+\bar x \\
      0 \\
      1+ y+\bar z + \bar y \bar z \\
      \end{pmatrix}$\\
    \hline
    \end{tabular}
    \label{tab:JW3Dsummary}
\end{table*}

\section{``Naive'' Jordan-Wigner in 2D}\label{app:naiveJW}
In Sec. \ref{ex:line2D_1}, the JW duality was considered for a fermionic system with vertical and horizontal line symmetries in 2D. Since the interaction terms can be chosen to commute, the dual symmetry is anomaly-free and allows us to further perform a KW duality on the spin system. The result of the combined duality is
\begin{align}
    i \gamma \gamma'  & \rightarrow X, \\
\begin{array}{cc}
  \gamma' & \gamma\\
 \gamma' &\gamma
\end{array}  & \rightarrow
\begin{array}{cc} 
 Z & Z\\
Z &Z
\end{array}.
\end{align}
This mapping can be obtained by a ``naive'' JW transformation in 2D as follows: We define a total ordering of sites row by row in a 2D square lattice as
\begin{align}
\begin{matrix}
\cdots&<& (i-1,j-1)  &<& (i,j-1)&<& (i+1,j-1) &<&\cdots \\
\cdots&<& (i-1,j)  &<& (i,j)&<& (i+1,j) &<&\cdots \\
\cdots&<& (i-1,j+1)  &<& (i,j+1)&<& (i+1,j+1) &<&\cdots
\end{matrix}
\end{align}
Then the Majorana operators can be mapped as
\begin{align}
\begin{matrix} 
    \gamma_{(i,j)} &\rightarrow& \displaystyle Z_{(i,j)} \prod_{(k,l)<(i,j)}   X_{(k,l)},\\
  \gamma_{(i,j)}' &\rightarrow& \displaystyle Y_{(i,j)} \prod_{(k,l)<(i,j)} X_{(k,l)},  \\
  P_{(i,j)} =- i  \gamma_{(i,j)}\gamma_{(i,j)'} &\rightarrow &X_{(i,j)},
  \end{matrix}
\end{align}
which reproduces the duality above.

The obvious problem of this map for dualizing a fermionic system with global symmetry is that bilinears of Majoranas between different rows get mapped to nonlocal spin operators. However, with fermion parity conservation on every horizontal line, such operators are forbidden, and so all symmetric operators under subsystem fermion parity get mapped to local spin operators.

\section{Non-Existence of Commuting interaction terms for Twisted Dual models}\label{app:anomalyproof}
This appendix is devoted to proving that for certain fermionic systems, one cannot redefine interaction terms $\bs S$ by attaching local fermion parities so that $\inner{\bs S,\bs S}_F=0$. The models where we are able to prove so are those with global symmetry in 2D and 3D, and those with Fibonacci symmetry in 2D and 3D.
\subsection{2D $\mathbb Z_2^F$ global symmetry}
Let us demonstrate with the simplest example where it is known that the dual theory is necessarily anomalous\cite{GaiottoKapustin2016}. From the model in Sec. \ref{ex:Global2D}, we have
\begin{align}
\inner{\bs P,\bs S} &= \begin{pmatrix} 1+x & 1+y\end{pmatrix},\\
 \inner{\bs S,\bs S}_F&=\begin{pmatrix} 0 & 1+\bar x y\\1+x\bar y &0 \end{pmatrix}.
\end{align}
For $i=j=1$, Eq. \eqref{eq:anomalyfree} reads
\begin{equation}
   (1+\bar x)  f_1 +  (1+x)\bar f_1 =0.
    \label{equ:2Dsolve1}
\end{equation}
where we have dropped the index $k$, since there is only one site per unit cell.

We wish to solve for solutions of such linear equation in $R=\mathbb F_2 [x^{\pm1},y^{\pm1}]$. A method of solving such equations is by first going to the fraction field $\text{Frac}(R)$, consisting of formal fractions of elements in an integral domain $R$. Formally,
\begin{equation}
\text{Frac}(R) = \frac{\left.\left \{ \frac{f}{g} \right |  f,g \in R, g \ne 0   \right \} }{ \left (\frac{f}{g} \sim \frac{f'}{g'} \ \text{if} \ fg'=f'g \in R \right ) }.
\end{equation}
The denominator denotes an equivalence class, where two fractions are equal if their cross-multiplications in $R$ are equal.

Since $\text{Frac}(R)$ is equipped with an involution, we can separate out its ``symmetric'' part. Let us define the symmetric subfield
\begin{equation}
    S =  \left. \left \{ f \in \text{Frac}(R) \right|  f =\bar f \right \}.
\end{equation}
It turns out that a single extra ``antisymmetric'' generator is sufficient to extend $S$ back to $\text{Frac}(R)$. Let us choose this generator to be $x$. To see why, we note that
\begin{align}
    \bar x &= (x+\bar x) + (1) x,\\
    x^2 &= (1) + (x+ \bar x) x,\\
    y &=  \left (\frac{\bar x y + \bar y x}{x+\bar x} \right) +  \left (\frac{y+\bar y}{x+\bar x} \right)x,
\end{align}
where the quantities in brackets are in $S$. All other monomials can be constructed recursively from these relations.

We now solve Eq. \eqref{equ:2Dsolve1}. First, we decompose $f_1$ into its symmetric and antisymmetric parts
\begin{equation}
    f_1 =f_1^s + f_1^a x,
\end{equation}
where $f_1^s, f_1^a \in S$.
Inserting this, we obtain
\begin{equation}
    (f_1^s + f_1^a) (x+\bar x) =0.
\end{equation}
Thus, we need $f_1^s = f_1^a$, meaning
\begin{equation}
    f_1 =  (1+x)f_1^s \ ; \ f_1^s \in S.
\end{equation}
Projecting solutions back to $R$, we require $f_1^s$ to be an element of $R$ that is invariant under the involution. An identical exercise shows that 
\begin{equation}
    f_2 =  (1+y)f_2^s \ ; \ f_2^s  = \bar f_2^s \in R.
\end{equation}
Let us now finally prove that there are no solutions to Eq. \eqref{eq:anomalyfree}. Its $(1,2)$ component reads
\begin{equation}
   (1+\bar x)  f_1 +  (1+y)\bar f_2 =1+\bar x y.
\end{equation}
Inserting the solutions, we find
\begin{equation}
   (1+\bar x)(1+x)  f_1^s +  (1+y)(1+\bar y) f_2^s =1+\bar x y.
\end{equation}
Since the left hand side is invariant under the involution, while the right hand side is not, there is no valid solution.

The proof is identical for the 3D case with global symmetry.

\subsection{Fibonacci fractal symmetry}
We show that the fermion model protected by the Fibonacci fractal symmetry in Secs. \ref{ex:Fibonacci2D} and \ref{ex:Fibonacci3D} cannot have commuting interaction terms. The interaction term is given by
\begin{align}
\bs S= \lc \begin{pmatrix}
x(1+y+\bar y)
\\
\hline 
1
\end{pmatrix}
\end{align}
so that
\begin{align}
\inner{\bs P,\bs S} &= 1+ x(1+y+\bar y), & \inner{\bs S,\bs S}_F=y^2+\bar y^2,
\end{align}
and Eq. \eqref{eq:anomalyfree} reads
\begin{align}
    (1+\bar x (1+y+\bar y))f +   (1+ x (1+y+\bar y))\bar f = y^2+\bar y^2.
\end{align}
Inserting $f= f^s +x f^a$, we find
\begin{align}
   (f^s(1+y+\bar y) +   f^a)(x+\bar x) = y^2+\bar y^2.
\end{align}
Although there exists solutions in $\text{Frac}(R)$, there are no valid solutions in $R$. This is because $(x+\bar x)$ does not have an inverse in $R$\footnote{In the case of finite but large system size $L$, an inverse can exist for certain polynomials. However, they are not local.}. 

The proof above also shows that the stabilizers of the twisted fractal code cannot be connected to the usual fractal code in the presence of translation symmetry. It would be interesting to see if their ground states are actually distinct or not.

\onecolumngrid
\section{Symmetry Action on the boundary of a fractal SPT}\label{app:symmetryaction}
In this appendix, we calculate the symmetry action on the boundary of the 3D fractal SPT. The symmetry acts on the lower half region
\begin{equation}
  \lc\begin{pmatrix}
    0\\
    \hline 
   \sum_{i}\sum_{j=-\infty}^0 \bar y^i(1+x+\bar x)^iz^j
    \end{pmatrix},
\end{equation}
and the stabilizer is given by
\begin{equation}
    \lc\bs H_{SPT}= \begin{pmatrix}
     x^2 + \bar x^2 +(y+ \bar y) (1+x+\bar x) +z (1+ x^2 +y(1+x+\bar x))+ \bar z(1+\bar x^2 + \bar y(1+x+\bar x)) \\
     \hline
   1
    \end{pmatrix}.
\end{equation}
Since the stabilizer acts on three consecutive planes, we can use the stabilizer to substitute
\begin{equation}
 \lc\begin{pmatrix}
    0\\
    \hline 1
    \end{pmatrix}
     \rightarrow \lc \begin{pmatrix}
     x^2 + \bar x^2 +(y+ \bar y) (1+x+\bar x) +z (1+ x^2 +y(1+x+\bar x))+ \bar z(1+\bar x^2 + \bar y(1+x+\bar x)) \\
     \hline
   0
    \end{pmatrix}.
\end{equation}
from the second layer downwards (i.e., for any $z^j$ where $j<0$). The symmetry is then
\begin{equation}
  \lc\begin{pmatrix}
    \sum_{i}\sum_{j=-\infty}^{-1} \bar y^i(1+x+\bar x)^iz^j \left [x^2 + \bar x^2 +(y+ \bar y) (1+x+\bar x) +z (1+ x^2 +y(1+x+\bar x))+ \bar z(1+\bar x^2 + \bar y(1+x+\bar x)) \right ]\\
    \hline 
    \sum_{i} \bar y^i(1+x+\bar x)^i
    \end{pmatrix}.
\end{equation}
The sum for the positions of the Pauli $Z$ above separated by layer (i.e. by the degree of $z$) is
\begin{align}
z^0&  \left [\sum_{i} \bar y^i(1+x+\bar x)^i \left ( 1+ \bar x^2 +  y (1+x+\bar x)   \right)  \right ] \nonumber  \\
+\bar z&\left [\sum_{i} \bar y^i(1+x+\bar x)^i \left (x^2 + \bar x^2 +(y+ \bar y) (1+x+\bar x) +  (1+ \bar x^2 + y (1+x+\bar x))  \right  )   \right ]\nonumber \\
+\sum_{j=2}^\infty \bar z^j& \left [\sum_{i} \bar y^i(1+x+\bar x)^i \left ( x^2 + \bar x^2 +(y+ \bar y) (1+x+\bar x) + (1+ x^2 +y(1+x+\bar x))+ (1+\bar x^2 + \bar y(1+x+\bar x)  \right  )   \right ]
\end{align}
The last line cancels completely, while for the other two lines we can shift $i$ to simplify the expression. This results in
\begin{equation}
  \lc\begin{pmatrix}
   x^2 + \bar z\bar x^2 \sum_{i} \bar y^i(1+x+\bar x)^i \\
    \hline 
   \sum_{i} \bar y^i(1+x+\bar x)^i
    \end{pmatrix}.
\end{equation}
There are remaining Pauli $Z$'s on the layer $\bar z$, but they do not anticommute with anything else in that layer. Therefore, the algebra of the symmetry operators matches that of
\begin{equation}
  \lc\begin{pmatrix}
   x^2 \sum_{i} \bar y^i(1+x+\bar x)^i \\
    \hline 
   \sum_{i} \bar y^i(1+x+\bar x)^i
    \end{pmatrix},
\end{equation}
which upon inversion is the anomalous symmetry of the 2D system given in Eq. \eqref{equ:2Dfractaldualsym} or Fig. \ref{fig:fractalsym}.

\twocolumngrid

\bibliography{references}

\end{document}